\documentclass[10pt,a4paper]{article}

\usepackage{amsmath}
\usepackage{amssymb}

\begin{document}

\newtheorem{lem}{Lemma}
\newtheorem{defin}{Definition}
\newtheorem{theor}{Theorem}
\newtheorem{rem}{Remark}
\newtheorem{prop}{Proposition}
\newtheorem{cor}{Corollary}
\newenvironment{demo}
{\bgroup\par\smallskip\noindent{\it Proof: }}{\rule{0.4em}{0.4em}
\egroup}
\newenvironment{ex}
{\bgroup\par\smallskip\noindent{\bf Example }}{\rule{0.5em}{0.5em}
\egroup}

\title{Discrete symmetries determining scalar quantum modes on the de Sitter spacetime}

\author{Ion I.  Cot\u{a}escu\thanks{E-mail:~~~cota@physics.uvt.ro} and 
Gabriel Pascu\thanks{E-mail:~~~gpascu@physics.uvt.ro}\\
{\small \it West University of Timi\c{s}oara,}\\
{\small \it V.  P\^{a}rvan Ave.  4, RO-300223 Timi\c{s}oara, Romania}}

\maketitle

\begin{abstract}
The action of the discrete symmetries on the scalar mode functions of the de Sitter spacetime is studied.
The invariance with respect to a combination of discrete symmetries is put forward as a criterion to select a certain vacuum out of a family of vacua. The implications of the choices for eigenfunctions of various common sets of commuting operators are explored, and the results are compared to the different original choices from the literature that have been utilized in order to exhibit thermal effects. 
\end{abstract}



\newpage

\section{Introduction}

An essential step in constructing a quantum field theory is constituted by the quantum mode functions. As is well known, in curved spacetimes there is an ambiguity in choosing quantum modes. This is sometimes used to discuss certain quantum effects and justify phenomena that may or may not have a basis in the physical reality. Therefore, the problem of selecting appropriate modes is of crucial importance.

Mathematically, the scalar  mode functions can be found by solving the Klein-Gordon equation in suitable local charts, allowing the separation of variables that introduces integration constants with some physical meaning. \cite{nachtmann,borner_durr,chernikov_tagirov,tagirov} Another possibility is to use  the standard  procedure of the quantum theory where the quantum modes are defined as common eigenstates of  several complete sets of commuting operators that must include the operator of the field equation. \cite{cota_crucean_pop} In general,  any set of commuting operators (s.c.o.) determines a representation regardless if it is complete or not.  These s.c.o. have to be chosen among the isometry generators which are conserved on the given manifold. In the case of the de Sitter one, which has $SO(1,4)$ isometries, we have at our disposition the large algebra freely generated by the $SO(1,4)$ generators. Sadly, this algebra has Abelian sub-algebras that are too small and as such we cannot pick complete sets of commuting operators like in the flat case - invariably one operator is missing. Thus any s.c.o. leaves two integration constants that cannot be determined as eigenvalues of an operator with precise physical meaning. This difficulty is avoided using supplemental physical criteria for selecting quantum modes, such that we face now with many types of quantum modes, corresponding to different vacua, between which the Bogolyubov transformations lay out various quantum effects.   

Under such circumstances, we may ask if the missing operator of a s.c.o. could be compensated by another symmetry which was not considered so far. Obviously, this can be only a discrete isometry of the $O(1,4)$ group. Some time ago, Allen pointed out that there is a representation of the mode functions defining the Euclidean vacuum where these are symmetric with respect to the charge conjugation $C$, combined with the antipodal  transformation \cite{allen}, denoted here by $T$. 
A similar result was obtained for de Sitter spacetimes with any number of space dimensions \cite{bousso}.

Thus, a condition involving certain discrete symmetries was shown to highlight this Euclidean vacuum, which also coincides with the Bunch-Davies vacuum \cite{bunch}. This condition is recognized as the $PCT$-invariance of the de Sitter mode functions \cite{bousso} and is the approach which we spotlight in this paper. It is based on the connection between the mode functions and the fundamental $PCT$ symmetry- which is not significant only on the Minkowski spacetime, but also on de Sitter.

Alternatively, there is another significant way by which one can introduce the Euclidean vacuum. It is the only de Sitter invariant vacuum for which short-distance singularities in the Hadamard function are of certain form- called the Hadamard form. This assures that the commutator two-point function behaves in the same way in the flat space-time limit, i.e. the singularities arising in the coincidence limit have the same form as in ﬂat Minkowski space. This property can be used as a definition for this {\it privileged} vacuum. For a modern detailed derivation on how to explicitly find the modes that can be obtained in this way, see Ref. \cite{joung}.

While there are strong indications that the physical vacuum satisfies the Hadamard condition \cite{fukuma} and the alpha-vacuum family are not suitable candidates for the free theory \cite{brunetti}, there is interest for them in the interacting field theory. For instance, Goldstein and Lowe give a thorough review of the objections, but in defense show that the $\alpha$-vacua do indeed have a well-defined perturbative expansion which gives finite renormalized amplitudes in the conventional way. \cite{goldstein} 

As a third method to fixing the Euclidean vacuum for modes which are momentum eigenfunctions, there is also the possibility of selecting the modes by taking a "rest frame" limit, where the positive and negative frequency eigenstates separate as in special relativity. \cite{cota_rest,cota_pascu_dregoesc} 

In general, there are a lot of different but equivalent parametrisations for the family of vacua. \cite{chernikov_tagirov,bousso,schomblond,mottola} Usually, the introduction of the vacua family is made starting from the Euclidean modes, via the MA transformation (shorthand for Mottola-Allen) which was actually used earlier in Ref. \cite{schomblond}, but it can also be made in reverse, starting from any mode functions. For example, in \cite{spradlin} the starting point is a different basis function set than the one corresponding to the Euclidean vacuum, and as such the transformation is denoted by MA$^\prime$, parametrisation which is to be regarded as inequivalent to the MA one from the point of the reference functions.

In this sense, the present work uses parametrisations similar to MA$^\prime$ (depending on a case by case basis), but without making use of any other supplemental condition (like the Hadamard one\cite{spradlin}) than invariance with respect to discrete symmetries.

With regards to  the quantum theory formalism, Mottola \cite{mottola} selects the different solutions to represent the {\it in} and {\it out} states of a process, and thus interprets the Bogolyubov transition coefficients as giving rise to a phenomenon of particle production. Bogolyubov coefficients between spherical modes in various representations have also been computed by Sato and Suzuki. \cite{sato}

On the other hand, the Hadamard function approach is relevant especially because of the implications for cosmological settings \cite{collins,deboer} or the dS/CFT correspondence \cite{bousso,spradlin}, the structure of the vacuum family being a perequisite of such studies. 

In the present paper we pursue the first mentioned line of reasoning, looking for the general discrete symmetries able to determine completely systems of mode functions globally defined as eigenfunctions of concrete s.c.o. To this end, besides the above mentioned discrete symmetries, we consider in addition the parity, $P$, and a special transformation, $N$, which changes the sign of an index concentrating the principal constants of the theory. In this approach  we can study exhaustively the effects of all the $O(1,4)$  discrete transformations or those combined with the $N$ one, obtaining new interesting results. First of all we  show  that there are only two global discrete symmetries, called here $PCN$ and $PCT$,  able to take over the role of the missing operator of any s.c.o. Consequently, we derive two types of bases (i.e., orthonormal and complete systems of mode functions) in any representation and any local chart, namely $PCN$-invariant bases and $PCT$-invariant ones. We prove that, in general, these form families labelled by continuous parameters whose properties depend on the value of mass -  above or below a certain characteristic mass threshold. Thus we are able to identify the bases of physical interest, finding that  above this threshold there is only one $PCN$-invariant basis and a $PCT$-invariant one which, in addition, is symmetric under the change of the sign of the mentioned index. This last basis corresponds to the Bunch-Davies or Euclidean vacuum as previously mentioned, while the $PCN$-invariant basis produces a vacuum state which is related to the Euclidean one through the Bogolyubov transformation whose coefficients suggest the standard thermal interpretation \cite{hawking}.  

We start in the second section by reviewing some features of the theory of scalar fields on curved manifolds focusing on the effects of the undefined inner product which separates positive and negative frequencies modes. We define the complete s.c.o. able to determine completely a basis, fixing its frequencies separation. We show that each well-defined basis deals with a special property we call here the closure condition. In the case of incomplete s.c.o., this condition is not naturally
fulfilled, but this can be replaced by an additional global discrete symmetry. In other words, an incomplete s.c.o. endowed with a discrete symmetry is able to determine a basis just like a complete s.c.o. All this procedure requires to have conserved operators for building s.c.o. and global discrete symmetries. 

In section \ref{role_of_isometries} we concentrate on the de Sitter symmetries, showing that the generators of the $SO(1,4)$ isometries give rise to a large algebra of conserved operators. Among them we can chose different s.c.o., each one determining its own representation.  Here we consider five representations, pointing out that their s.c.o. are incomplete, so that we must resort to discrete transformations able to replace the closure condition. A rapid inspection points to the global de Sitter discrete symmetries, $P$ and $T$,  which can be combined with the above  mentioned $C$ and $N$ transformations giving the $PCT$ and $PCN$  symmetries defining specific bases and corresponding vacuum states.    

The next section is devoted to the concrete examples of the momentum and energy representations that can be studied in the conformal chart of the de Sitter manifold. We define the general bases of these representations that depend on two arbitrary integration constants and study their behaviour under discrete symmetries. Thus we find that there are only $PCT$ and $PCN$ invariant bases, since the $PC$ symmetry selects only null mode functions which have no physical meaning. Finally, we identify  the Bogolyubov transformation  generating the usual thermal coefficients in both the representations under consideration in this section.  

In sections \ref{r34} and \ref{r5} we study the spherical modes. While the previous section is discussed thoroughly in order to provide a rigorous template, the same considerations are punctually considered for the spherical modes, pointing out the differences that arise. At last, the global representation featured in the works of Mottola and Allen is analysed in this framework, illustrating the agreement with our proposed criterion of selecting the Euclidean vacuum.

\section{Scalar quantum modes}

The scalar fields $\Phi: M\to {\Bbb C}$ of mass $m$, minimally coupled to the gravity of a curved manifold $(M,g)$,  satisfy the Klein Gordon equation ${\cal E}\Phi=m^2\Phi$ given by the Klein-Gordon operator 
\begin{equation}\label{KG}
{\cal E}=-\frac{1}{\sqrt{g}}\,\partial_{\mu} \left( \sqrt{g}\,
g^{\mu\nu}\partial_{\nu} \right)\,,\quad g=|{\rm det} g_{\mu\nu}|\,,
\end{equation}
which determines different quantum modes. In general, the mode functions $f\in{\cal F}$, behave as tempered distributions or square integrable functions with respect to the undefined  inner (or scalar) product 
\begin{equation}\label{SPgen}
\langle f,f'\rangle=i\int_{\Sigma} d\sigma^{\mu}\sqrt{g}\left( f^*
\partial_{\mu} f' -f'\partial_{\mu}f^*\right)=i\int_{\Sigma} d\sigma^{\mu}\sqrt{g}\,  f^*\stackrel{\leftrightarrow}{\partial}_{\mu}f' \, \in {\Bbb C} \,.
\end{equation}
The square integrable functions $f\in {\cal H}\subset{\cal F}$ may have 'squared norms' $\langle f,f\rangle$ of any sign,  splitting  thus  the space ${\cal F}$ as
\begin{equation}
f \in \left\{ 
\begin{array}{lll}
{\cal H}_+\subset{\cal F}_+& {\rm if}& \langle f,f\rangle>0\,,\\
{\cal H}_0\subset{\cal F}_0& {\rm if}& \langle f,f\rangle=0\,,\\
{\cal H}_-\subset{\cal F}_-& {\rm if}& \langle f,f\rangle<0\,.\\
\end{array}
\right.
\end{equation} 
Physically speaking, the mode functions of  ${\cal F}_{\pm}$ are of positive/negative frequencies, while those of  ${\cal F}_0$ do not have a physical meaning. For any $f\in {\cal F}_+$ we have $f^*\in {\cal F}_-$ so that $\langle f^*, f^*\rangle =-\langle f, f\rangle$, but whether  $f^*=f$ then  $f\in {\cal  F}_0$, since $\langle f, f\rangle=0$.   In fact,  ${\cal H}$ is a Krein space while ${\cal F}_{\pm}$ are the spaces of  tempered distributions of the Hilbertian triads associated to the Hilbert spaces ${\cal H}_{\pm}$ (equipped with the scalar products $\pm \langle~,~\rangle$). 
\begin{defin}
 A complete system of orthonormal mode functions, $\{f_{\alpha}\}_{\alpha\in I}\subset{\cal F}_+$ forms a basis of positive frequencies in ${\cal F}_+$ related to the negative frequencies one, $\{f^*_{\alpha}\}_{\alpha\in I}\subset{\cal F}_-$. 
\end{defin} 
In this way one defines a  frequencies separation  associated to  a specific vacuum state of the Fock space. It is known that  two different bases define different vacuum states when these are related among themselves through a non-trivial Bogolyubov transformation that mixes the positive and negative frequency modes. Otherwise the vacuum state remains the same.
\begin{defin}\label{equiv}
Two different bases related through a trivial  Bogolyubov transformation  are equivalent, corresponding to the same vacuum state.
\end{defin}   
  
 In general,  a basis in ${\cal F}_+$ can be introduced as  a system of common eigenfunctions of a complete s.c.o.,  formed by hermitian operators $A$, obeying   $\langle f',Af\rangle=\langle Af',f\rangle$.  Then the eigenfunctions are orthogonal, corresponding to real-valued eigenvalues.  
 \begin{lem}
 Given the hermitian operator $A:{\cal F}\to {\cal F}$ and the eigenvalues problem $A f_{\alpha}=\alpha f_{\alpha}$  with $f_{\alpha}\in {\cal F}_+$  (and $\alpha\in{\Bbb R}$), then  $f^*_{\alpha}\in{\cal F}_-$  are eigenfunctions of the same operator, $A f^*_{\alpha}=\bar{\alpha} f^*_{\alpha}$, where 
$\bar{\alpha}=\alpha$ if $A^*=A$ or $\bar{\alpha}=-\alpha$  if  $A^*=-A$.
\end{lem}
Hereby we see that the eigenfunctions $f_{\alpha}$ and $f^*_{\bar{\alpha}}$ correspond to the same eigenvalue $\alpha$ giving rise to a supplemental  degeneracy as long as $f_{\alpha}$ and $f^*_{\bar{\alpha}}$ are linearly independent. Obviously, this degeneracy has to be removed when the system of mode functions is well-defined by a complete s.c.o.   
\begin{defin}\label{complete}
 A system of commuting operators $\{A_1,A_2,...\}$ is called complete if  all its common eigenfunctions that satisfy 
\begin{equation} 
A_af_{\alpha_1,\alpha_2,...}=\alpha_a f_{\alpha_1,\alpha_2,...}\,,\quad  a=1,2,..., 
\end{equation}
accomplish the closure condition
\begin{equation}\label{fetaf}
f^*_{\bar{\alpha}_1,\bar{\alpha}_2,...}=\eta f_{\alpha_1,\alpha_2,...}\,,
\end{equation} where $\eta$ is a phase factor. 
\end{defin}
This condition can be seen as a {\em discrete symmetry} due to some  discrete transformations which change $\alpha\to \bar{\alpha}$ combined with  the charge conjugation $f\to Cf=f^*$. 
\begin{ex}{\bf 1:  Minkowskian plane waves}
The plane waves in  Cartesian coordinates $\{x\}$ of the Minkowski spacetime  are the eigenfunctions $f_{\bf p}$ of the complete s.c.o. $\{{\cal E},P_{\mu}=-i\partial_{\mu}\}$  corresponding to  the eigenvalues $\{m^2, p_{\mu}\}$. These satisfy the condition $f^*_{-{\bf p}}=f_{\bf p}$, which relates to the universal $PCT$ invariance of special relativity  
\end{ex}
\begin{rem}
When the s.c.o. is incomplete, the closure condition (\ref{fetaf}) is not naturally  accomplished, but it can be replaced  by an additional discrete symmetry.  
\end{rem} 
In what follows, we would like to elaborate on this matter in the particular case of the de Sitter manifold.

\section{The role of the de Sitter isometries} \label{role_of_isometries}

Our main purpose is to construct covariant fields, globally defined on the entire de Sitter manifold, exploiting its  global $O(1,4)$ symmetries. The  fields defined on $(M,g)$ transform under continuous isometries according to the covariant representations of the group $SO(1,4)$, generated by conserved operators that form  covariant representations of the $so(1,4)$ algebra. Among these operators, which commute  with the those of the field equations, we can choose suitable s.c.o. determining  bases as complete  systems of common eigenfunctions. However, as we shall see below,  there are no complete s.c.o., and thus we must resort to discrete symmetries in order to complete the definitions of the bases.   

\subsection{Conserved observables} \label{3_1}

Let $(M,g)$ be  the de Sitter spacetime  defined as the hyperboloid of radius $1/\omega$ \footnote{We denote by $\omega$ the Hubble de Sitter constant, since  $H$ is reserved for the energy operator} in the five-dimensional flat spacetime $(M^5,\eta^5)$ of coordinates $z^A$  (labelled by the indices $A,\,B,...= 0,1,2,3,4$) and metric $\eta^5={\rm diag}(1,-1,-1,-1,-1)$. The local charts $\{x\}$  can be introduced on $(M,g)$ giving the set of functions $z^A(x)$ which solve the hyperboloid equation,
\begin{equation}\label{hip}
\eta^5_{AB}z^A(x) z^B(x)=-\frac{1}{\omega^2}\,.
\end{equation}
Here we use the chart $\{t,{\bf x}\}$ of the comoving frame with the conformal time $t$ and Cartesian spaces coordinates $x^i$ ($i,j,...=1,2,3$) defined by
\begin{eqnarray}
z^0(x)&=&-\frac{1}{2\omega^2 t}\left[1-\omega^2({t}^2 -{\bf x}^2)\right]\,,
\nonumber\\
z^i(x)&=&-\frac{x^i}{\omega t} \,, \label{Zx}\\
z^4(x)&=&-\frac{1}{2\omega^2 t}\left[1+\omega^2({t}^2 - {\bf x}^2)\right]\,.
\nonumber
\end{eqnarray}
This chart  covers the expanding part of $M$ for $t \in (-\infty,0)$ and ${\bf x}\in {\Bbb R}^3$, while the collapsing part is covered by a similar chart with $t >0$. Both these charts have the conformal flat line element
\begin{equation}\label{mrw}
ds^2=g_{\mu\nu}(x)dx^{\mu}dx^{\nu}=\frac{1}{\omega^2t^2}(dt^2- d{\bf x}\cdot d{\bf x})\,.
\end{equation}

The scalar field $\Phi: (M,g)\to {\Bbb C}$  transforms under the de Sitter isometries $x\to x'=\phi(x)$, according to the natural representation $\Phi\to \Phi\circ\phi^{-1}$,  generated by genuine orbital generators. In the standard parametrization with skew-symmetric real parameters, the basis-generators of this representation $X_{(AB)}=-i k_{(AB)}^{\mu}\partial_{\mu}$, are produced  by the Killing vectors $k_{(AB)}$ associated to the group parameters. \cite{cota2011} Here, it is convenient to consider the $so(1,4)$ basis constituted by the energy $H$, momentum $P_i$, angular momentum $J_i$ and the supplemental Abelian generators $Q_i$. In  the conformal chart, these basis-generators read 
\begin{eqnarray}
H &\equiv &  \omega X_{(04)}=-i\omega(t\,\partial_t+ {x}^i {\partial}_i)\,,\label{Ham}\\
P_i&\equiv & \omega(X_{(i0)}-X_{(i4)})=-i\partial_i\,,\label{Pi}\\
J_i&\equiv & \textstyle{\frac{1}{2}}\varepsilon_{ijk}X_{(jk)}=-i\varepsilon_{ijk}x^j\partial_k\,,\\
Q_i &\equiv & \omega(X_{(i0)}+X_{(i4)})=-2i \omega^2 x^i H+
\omega^2({\bf x}^2-t^2)P_i \,.\label{Qi}
\end{eqnarray}
Other important generators are those of the Lorentz boosts $K_i=X_{(0i)}$, and the Runge-Lenz type ones, $R_i=X_{(i4)}$, which allow us to construct the first  Casimir operators, $C^1_{so(1,3)}={\bf J}^2-{\bf K}^2$  and    $C^1_{so(4)}={\bf J}^2+{\bf R}^2$,  of the principal subgroups, $SO(1,3)$ and respectively $SO(4)$.  All these operators represent conserved observables, since they commute with the Klein-Gordon one,  ${\cal E}=C^1_{so(1,4)}$, which is just the first  Casimir operator of the group $SO(1,4)$. We remind the reader that in the natural representation, all the second Casimir operators of  the mentioned groups  vanish, since there is no spin.  

In practice, it is convenient to replace the Klein-Gordon operator ${\cal E}$ by the new operator 
\begin{equation}\label{OKGN}
{\cal N}=\frac{1}{\omega^2}\,{\cal E}-\frac{9}{4}= -t^2\left(\partial_t -\frac{1}{t}\right)^2 +t^2\Delta-\frac{1}{4}\,,
\end{equation} 
so that  the Klein-Gordon equation becomes 
\begin{equation}\label{KGN}
{\cal N}\Phi=\left(\mu^2-\lambda^2\right) \Phi\,,\quad  \mu=\frac{m}{\omega}\,,
\end{equation}
where $\lambda$ depends on the coupling to gravity, e.g. $\lambda=\frac{3}{2}$ for the minimal coupling and $\lambda=\frac{1}{2}$ for the conformal one.   
\begin{rem}
The solutions of the Klein-Gordon equation behave differently in the domains $\mu>\lambda$ and $\mu<\lambda$, which have to be considered separately, but using the unique real-valued  parameter $\nu=\sqrt{|\mu^2-\lambda^2|}$. 
\end{rem}

The quantum theory is based on  the operator algebra freely generated by the basis-generators (\ref{Ham})-(\ref{Qi})  that can be extended even with non-differential conserved operators, related to the differential ones.  For example, we shall use  the operator of the momentum direction $\hat{\bf P}={\bf P}/({\bf P}\cdot{\bf P})^{\frac{1}{2}}$ that commutes with $H$. All these operators  are hermitian with respect to the scalar product (\ref{SPgen}) that in the chart $\{t,{\bf x}\}$ reads
\begin{equation}\label{SP2}
\langle f,f'\rangle=i\int \frac{d^3 x}{\omega^2 t^2} \,f ^*(t,{\bf x})
\stackrel{\leftrightarrow}{\partial_{t}}f'(t,{\bf x})\,.
\end{equation}

Now we can  consider different s.c.o., understanding that each s.c.o. defines its own  specific {\em representation} (R)  as in the  next table: \\

\begin{tabular}{llll}
&Set of commuting  & Representation (R)&Refs. \\ 
&operators (s.c.o.)&&\\
I.&${\cal N}, P_1,P_2,P_3$& momentum& \cite{nachtmann,borner_durr} \\ 
II.&${\cal N}, H, \hat{P}_1,\hat{P}_2,\hat{P}_3$ & energy& \cite{cota_crucean_pop}\\ 
III.&${\cal N}, {\bf P}^2, {\bf J}^2, J_3$ & momentum-angular momentum &  \cite{sato} \\
IV.&${\cal N}, H, {\bf J}^2, J_3$ & energy-angular momentum& \cite{pascu}\\ 
V.&${\cal N},C^1_{so(4)} , {\bf J}^2, J_3$ &unitary global & \cite{chernikov_tagirov,tagirov}\\ 
VI.&${\cal N},C^1_{so(1,3)} , {\bf J}^2, J_3$ & hyperbolic & \cite{sasaki,pfautsch}
\end{tabular}\\

\noindent We observe that  the s.c.o. I, defining  the mode functions of the  momentum representation R.I, cannot be {\em completed} as it happens in the flat case, since the energy operator  (\ref{Ham}) is not diagonal in this basis,  $[H,P_i]=i\omega P_i$.  Moreover, analysing the  others  s.c.o. we arrive at the conclusion: 
\begin{rem}
On the de Sitter spacetime one cannot extract complete sets of commuting  operators from  the whole algebra of conserved observables. 
\end{rem} 
This means that in the de Sitter geometry the systems of quantum modes can be globally defined, but only up to some arbitrary integration constants which may be specified  using some additional physical  criteria exhaustively discussed in the literature.

However, in what follows we concentrate only on the global discrete symmetries that help us to complete the global definitions of the systems of quantum modes.       

\subsection{Discrete symmetries}

In general, the local discrete symmetries changing the signs of several local coordinates are useless on curved backgrounds, since these depend on the concrete coordonatization.  Fortunately, in the de Sitter case  there exist {\em  global} discrete symmetries, related to the  $O(1,4)$ discrete transformations on $(M^5,\eta^5)$. The simplest ones, denoted by ${\pi_{[A]}}$, are those changing the sign of a single coordinate of  $M^5$, $z^A\to -z^A$. These transformations give rise to the discrete isometries $x\to x'=\phi_{[A]}(x)$  defined as $z[\phi_{[A]}(x)]=\pi_{[A]}[z(x)]$, which  satisfy $\phi_{[A]}\circ\phi_{[A]}=id$. We observe that the  isometries  produced by $\phi_{[0]}$ and  $\phi_{[4]}$ cannot be used for our purpose, since these are point-dependent \cite{cota2013}. Thus we remain with the isometries, $\phi_{[i]}$, which are simple mirror transformations of the space coordinates $x^i$, so that the (space) parity reads $\phi_{({\bf x})}=\phi_{[1]}\circ\phi_{[2]}\circ\phi_{[3]}$.  Another remarkable discrete transformation is the antipodal one, $z\to -z$, giving rise to the isometry $\phi_{(t)}=\phi_{(x)}\circ\phi_{({\bf x})}$ that plays the role of the time reversal in the conformal chart of  $(M,g)$ changing $t\to -t$. Notice that this transformation changes the expanding and collapsing portions of the de Sitter manifold between themselves. The action of the discrete isometries in spherical coordinates is briefly presented in  Appendix A.
\begin{rem}
The above discrete transformations are globally defined, independent on the local coordinates. Their action upon the scalar fields, $\Phi : M\to {\Bbb C}$,  can be written in any chart as $ \Phi \to\Phi'= \Phi\circ \phi_{[A]}$. 
\end{rem} 
In particular, we denote the parity by $ P=\phi_{({\bf x})}$ and the antipodal transformation by $T=\phi_{(t)}$. Then, in the conformal chart  $\{t,{\bf x}\}$ we can write
\begin{equation}
(P\Phi)(t,{\bf x})=\Phi(t,-{\bf x})\,, \quad  (T\Phi)(t,{\bf x})=\Phi(-t,{\bf x})\,.
\end{equation}  
In addition, we consider the charge conjugation $C$ giving  $C\Phi=\Phi^*$ and denote by $I$ the identity transformation.    
\begin{theor}
The discrete transformations $P$, $C$ and $T$ have the following properties
\begin{eqnarray}
P:{\cal F}_{\pm}\to{\cal F}_{\pm}&\quad& P=P^{\dagger}=P^{-1}\,,\\
C:{\cal F}_{\pm}\to{\cal F}_{\mp}&\quad& C=-C^{\dagger}=C^{-1}\,,\\
T:{\cal F}_{\pm}\to{\cal F}_{\mp}&\quad& T=-T^{\dagger}=T^{-1}\,.
\end{eqnarray}
\end{theor}
\begin{demo}
We observe first that all these transformations are involutions: $P^2=C^2=T^2=I$. Then we  calculate $\langle Pf,Pf'\rangle=\langle f,f'\rangle$, $\langle Cf,Cf'\rangle=-\langle f,f'\rangle$ and   $\langle Tf,Tf'\rangle=-\langle f,f'\rangle$ using the scalar product (\ref{SP2}) \end{demo}

A special discrete symmetry can be introduced in the de Sitter case since the Klein-Gordon equation (\ref{KGN})  allows simultaneously the particular solutions $f_{\nu}$ and $f_{-\nu}$, depending on the signed parameter $\pm \nu$. This conjuncture generates a new discrete symmetry we call here  $\nu$-symmetry, introducing the operator $N$  so that $Nf_{\nu}=f_{-\nu}$. This operator satisfies $N^2=I$ and  $N=-N^{\dagger}$, but this last property can be pointed out only on the concrete form of the mode functions.

The action of the discrete symmetries upon the basis-generators is quite simple if we observe that $T$ and $N$ conserve the form of all the isometry generators while $C$ changes the signs of all of them. It remains the parity $P$, which changes the signs of the generators $X_{(AB)}$ carrying only one space index. Thus we find that
\begin{equation}\label{HJPQ}
\left\{PC, H\right\}=0\,,\quad \left\{PC, J_i\right\}=0\,,\quad \left[PC, P_i\right]=0\,,\quad \left[PC, Q_i\right]=0\,.
\end{equation}
The  Casimir operators are $PC$ invariant since they are quadratic forms. We note that $P$ commutes with the other discrete transformations, but we cannot establish general commutation or anti-commutation rules among $C$, $T$ and $N$. 

With these ingredients, we are able to define the unitary transformations $PCT$ and $PCN$, which obey $(PCT)^{\dagger}=(PCT)^{-1}$ and  $(PCN)^{\dagger}=(PCN)^{-1}$. These discrete symmetries can be used as closure conditions for any  representation determined by an incomplete s.c.o. $\{{\cal N},A_1,A_2,A_3\}$, whose eigenvalues are denoted by $\{\mu^2-\lambda^2,\alpha_1,\alpha_2,\alpha_3\}$.
\begin{defin} \label{d4}
Given a complete system of common eigenfunctions $f_{\nu,\alpha_1,\alpha_2,\alpha_3}\in {\cal F}_+$ of the s.c.o.  $\{{\cal N},A_1,A_2,A_3\}$, we say that this forms\\
a) either a $PCT$-invariant basis (or, simply, $PCT$-basis)  obeying
\begin{equation}
PCT f_{\nu,\alpha_1,\alpha_2,\alpha_3}=\eta f_{\nu,\alpha'_1,\alpha'_2,\alpha'_3}\,,\quad  |\eta|=1\,,
\end{equation}
b) or a $PCN$-basis when 
\begin{equation}
PCN f_{\nu,\alpha_1,\alpha_2,\alpha_3}=\eta' f_{-\nu, \alpha'_1,\alpha'_2,\alpha'_3}\,,\quad  |\eta'|=1\,,
\end{equation}
where 
\begin{equation}
\alpha'_a = \left\{ \begin{array}{lll}
\alpha_a&{\rm if}&~ [PC,A_a]=0\,,\\
-\alpha_a&{\rm if}&\,\{PC,A_a\}=0\,,
\end{array}\right. \quad a=1,2,3\,.
\end{equation}
\end{defin}
Obviously, the corresponding  basis of negative frequencies are also $PCT$-invariant or $PCN$-invariant. 
In what follows, we shall see  that these are the only discrete symmetries able to play the role of the closure condition
(\ref{fetaf}). 

\section{Plane waves in R.I and R.II} \label{r12}

The eigenvalues problems of R.I and R.II  are related to each other, since the action of  the (non-differential) operator $\hat{\bf P}$ can be defined only by giving its spectral representation in R.I. The eigenfunctions of these representations are plane wave functions which can be easily studied in the conformal  chart $\{t,{\bf x}\}$, where we have the operators of section \ref{3_1} and  the inner product (\ref{SP2}).

\subsection{Equivalent bases of  R.I and R.II}

Let us consider first a  set of particular  eigenfunctions  $u_{\nu,{\bf p}}\in{\cal F}_+$ of the incomplete s.c.o. I, corresponding to the eigenvalues  $\{\mu^2-\lambda^2, p_1,p_2,p_3\}$.  This set  forms a basis of R.I if  its mode functions are  normalized in the momentum scale,
\begin{eqnarray}
\langle  u_{\nu,{\bf p}},u_{\nu,{\bf p}'}\rangle=-\langle  u_{\nu,{\bf p}}^*,u_{\nu,{\bf
p}'}^*\rangle&=&\delta^3({\bf p}-{\bf p}')\,,\label{ortou1}\\
\langle u_{\nu,{\bf p}},u_{\nu,{\bf p}'}^*\rangle&=&0\,,\label{ortou2}
\end{eqnarray}
obeying the completeness condition
\begin{equation}\label{comp}
i\int d^3p\,  u^*_{\nu,{\bf p}}(t,{\bf x}) \stackrel{\leftrightarrow}{\partial_{t}}
u_{\nu,{\bf p}}(t,{\bf x}')=(\omega t)^2\delta^3({\bf x}-{\bf x}')\,.
\end{equation}

According to our previous results \cite{cota_crucean_pop},  each basis $\{ u_{\nu,{\bf p}}\}$ of R.I can be related to an equivalent  one of  R.II  (in the sense of definition (\ref{equiv}))  through a trivial Bogolyubov  transformation. This is formed by the eigenfunctions $u_{\nu,E,{\bf n}}\in {\cal F}_+$ of the s.c.o. II,  corresponding to the eigenvalues $\{\mu^2-\lambda^2,E,n_1,n_2,n_3\}$ where $E\in {\Bbb R}$ and ${\bf n}\in S^2$. These mode functions are normalized in the energy scale \cite{cota_crucean_pop},
\begin{eqnarray}
\langle u_{\nu,E,{\bf n}},u_{\nu,E',{\bf n}^{\,\prime}}\rangle=- \langle u^*_{\nu,E,{\bf
n}},u^*_{\nu,E',{\bf n}^{\,\prime}}\rangle&=& \delta(E-E')\,\delta^2 ({\bf n}-{\bf
n}^{\,\prime})\,,\label{ortou11}\\
\langle u_{\nu,E,{\bf n}},u^*_{\nu,E',{\bf n}^{\,\prime}}\rangle&=&0\,, \label{ortou12}
\end{eqnarray}
satisfying the completeness condition
\begin{equation}\label{comp2}
i\int_0^{\infty}dE\int_{S^2} d\Omega_{\bf n} \left [u^*_{\nu,E,{\bf n}}(t,{\bf
x})\stackrel{\leftrightarrow}{\partial_{t}} u_{\nu,E,{\bf n}}(t,{\bf x}')
\right] =(\omega t)^2\delta^3 ({\bf x}-{\bf x}^{\,\prime})\,.
\end{equation}

In general, the bases  of R.I or R.II  are determined up to a pair of remaining  integration constants  $c_1$ and $c_2\in{\Bbb C}$ which cannot be determined by the eigenvalues problems. Therefore, we have to consider general  eigenfunctions of s.c.o. I  of the form 
\begin{equation}\label{modeF}
f_{(c_1,c_2)\,\nu,{\bf p}}=c_1 u_{\nu,{\bf p}}+c_2 u^*_{\nu,{-{\bf p}}}\,,\quad c_1,\, c_2\in {\Bbb C}\,,
\end{equation} 
and  equivalent eigenfunctions of  the s.c.o. II,
\begin{equation}\label{modeFE}
f_{(c_1,c_2)\,\nu, E,{\bf n}}=c_1 u_{\nu,E,{\bf n}}+c_2 u^*_{\nu,-E,{-{\bf n}}}\,.
\end{equation}

These mode functions can be organized manipulating the vectors  $c=(c_1 , c_2)$ of the space ${\Bbb C}^2$  endowed with the inner product defined as 
\begin{equation}\label{cc}
\langle c \cdot c'\rangle=c^*_1 c_1' - c^*_2 c_2'.
\end{equation}
This gives the forms $\langle  c\cdot c\rangle \in{\Bbb R}$  the role of squared norms and splits the space ${\Bbb C}^2$ in subspaces of  positive, negative or null vectors. For any positive or negative vector,  $c=(c_1,c_2)$  with $\langle c\cdot c\rangle\not=0$,  one can define the vector   $\bar{c}=(c^T)^*=(c^*_2 ,c^*_1)$, orthogonal to $c$, which obeys  $\langle c\cdot \bar{c}\rangle=0$ and  $\langle \bar{c}\cdot \bar{c}\rangle=-\langle c\cdot c \rangle$. When   $c$  is a positive unit vector obeying $\langle c \cdot c\rangle=1$,  then $\bar{c}$  is a negative unit vector,  $\langle \bar{c} \cdot\bar{c}\rangle=-1$. The pair $\{c,\bar{c}\}$ forms a pseudo-orthonormal basis of  ${\Bbb C}^2$ in which any vector $c'$ can be written as $c'=c\langle c\cdot c'\rangle - \bar{c}\langle \bar{c}\cdot c'\rangle$ so that  
\begin{equation}\label{ccc} 
\langle c'\cdot c'\rangle=| \langle c'\cdot c\rangle|^2-| \langle c'\cdot\bar{ c}\rangle|^2\,.    
\end{equation}

With these preparations we can write the simple formulas of the scalar products of  arbitrary mode functions of R.I or R.II,
\begin{eqnarray}
\langle f_{c\,\nu,{\bf p}}, f_{c'\,\nu,{\bf p}'}\rangle &=&\langle c \cdot c'\rangle\, \delta^3({\bf p}-{\bf p}')\,,\label{scalarFF}\\
\langle f_{c\,\nu,E,{\bf n}},  f_{c'\,\nu,E',{\bf p}'}\rangle &=&\langle c\cdot c'\rangle\,\delta(E-E')\,\delta^2({\bf n}-{\bf n}')\,,\label{scalarFFE}
\end{eqnarray}
the integral
\begin{equation}\label{completFF}
i\int d^3p f_{c\,\nu,{\bf p}}^*(t,{\bf x})\stackrel{\leftrightarrow}{\partial_{t}} f_{c'\,\nu,{\bf p}}(t,{\bf x}')= \langle c \cdot c'\rangle\omega^2 t^2\delta^3({\bf x}-{\bf x}')\,,
\end{equation} 
and a similar one for R.II.  Hereby we understand that  different bases of R.I or R.II can be defined by choosing  pairs of non-null orthogonal  vectors, $c$ and $\bar{c}$,  for separating the positive and  negative frequencies. Taking  $c'=c$ or $c'=\bar{c}$ in Eqs. (\ref{scalarFF}), (\ref{scalarFFE})  and  (\ref{completFF})  we conclude that these bases are orthonormal and complete.  Moreover, following the method of Ref. \cite{cota_crucean_pop} we find that the bases with the same vector $c$ of R.I and R.II are equivalent when the starting ones,  $\{ u_{\nu,{\bf p}}\}$  and  $\{ u_{\nu,E,{\bf n}}\}$,  have this property. Therefore, we have reason to define:
\begin{defin}
For any positive unit vector $c\in{\Bbb C}^2$ (with $ \langle c \cdot c\rangle =1$) we  say that: \\
a) in R.I,  the set  
$\{f_{c\,\nu,{\bf p}}| {\bf p}\in {\Bbb R}^3_p\}\subset {\cal F}_+$ represents  the positive-frequency $c$-basis  while the set   
$\{f_{\bar{c}\,\nu,{\bf p}}| {\bf p} \in {\Bbb R}^3_p\}\subset {\cal F}_-$ forms the  negative-frequency one;\\
b) the  equivalent  $c$-bases of R.II are  $\{f_{c\,\nu,E,{\bf n}} | E\in {\Bbb R},{\bf n}\in S^2\}\subset {\cal F}_+$ and $\{f_{\bar{c}\,\nu, E,{\bf n}} | E\in {\Bbb R}, {\bf n}\in S^2\}\subset {\cal F}_-$.
 \end{defin}

Using such bases, the massive and charged scalar field can be expanded  in  R.I and R.II, 
\begin{eqnarray}
\Phi(t,{\bf x})&=&\int d^3p \left[ f_{c\,\nu,{\bf p}}(t,{\bf x}) a_c({\bf p})+ f^*_{c\,\nu,{\bf p}}(t,{\bf x}) b_c^{\dagger}({\bf p})\right]\\
&=&\!\!\int_{0}^{\infty}\!dE\int_{S^2} d\Omega_{\bf n} \left[ f_{c\,\nu,E,{\bf n}}(t,{\bf x}) a_c(E,{\bf n})+ f^*_{c\,\nu,E,{\bf n}}(t,{\bf x}) b_c^{\dagger}(E,{\bf n})\right]\!\!\,,
\end{eqnarray} 
in terms of particle ($a_c\,, a_c^{\dagger}$) and antiparticle ($b_c\,, b_c^{\dagger}$) operators corresponding to the frequency separation due to  these $c$-bases.  We note that the second integral is taken only over the positive energies, even though the mode functions in ${\cal F}_+$ are defined  for  any $E\in {\Bbb R}$. This is because we desire to use similar integration domains, ${\Bbb R}^3_p\sim S^2\times {\Bbb R}^+$, for these equivalent  representations. 

When one starts with the field $\Phi$, one can recover the field operators  using the inversion formulas
\begin{eqnarray}
&a_c({\bf p})=\langle f_{c\,\nu,{\bf p}}, \Phi \rangle\,,\quad & a_c(E,{\bf n})=\langle f_{c\,\nu,E,{\bf n}}, \Phi \rangle\,, \label{invers}\\
&b_c({\bf p})=\langle f_{c\,\nu,{\bf p}}, \Phi^{\dagger} \rangle\,,\quad & b_c(E,{\bf n})=\langle f_{c\,\nu,E,{\bf n}}, \Phi^{\dagger} \rangle\,,
\end{eqnarray}
bearing in mind that the field operators of R.I and R.II are related among themselves as in Ref. \cite{cota_crucean_pop}.
The canonical quantization of the field $\Phi$ requires  these field operators to satisfy the non-vanishing commutation relations 
\begin{eqnarray}
&&\left[a_c({\bf p}),a^{\dagger}_c({\bf p}')\right]=\left[b_c({\bf p}),b^{\dagger}_c({\bf p}')\right]=\delta^3({\bf p}-{\bf p}')\,,\label{comab}\\
&&\left[a_c(E,{\bf n}),a^{\dagger}_c(E',{\bf n}')\right]=\left[b_c(E,{\bf n}),b^{\dagger}_c(E',{\bf n}')\right]=\delta(E-E')\delta^2({\bf n}-{\bf n}')\,,\label{comabE}
\end{eqnarray}
for all the positive unit vectors  $c\in{\Bbb C}^2$. These properties are conserved by  the Bogolyubov transformations between different $c$-bases. 
 \begin{theor}
The field operators of  two different $c$-bases of R.I defined by the positive  unit vectors $c$ and $c\not=c'$  are related through the Bogolyubov transformation
\begin{eqnarray}
a_{c'}({\bf p})&=&\langle c'\cdot c\rangle a_c({\bf p}) +\langle c'\cdot \bar{c}\rangle b^{\dagger}_c(-{\bf p})\,,\\
b_{c'}({\bf p})&=&\langle c'\cdot c\rangle b_c({\bf p}) +\langle c'\cdot \bar{c}\rangle a^{\dagger}_c(-{\bf p})\,,
\end{eqnarray}   
whose coefficients satisfy
\begin{equation}\label{Bogo}
| \langle c'\cdot c\rangle|^2-| \langle c'\cdot\bar{ c}\rangle|^2=1\,.
\end{equation}
\end{theor}
\begin{demo} We use the expansions in both these bases, the inversion formulas (\ref{invers}) and Eq. (\ref{scalarFF}).
Eq. (\ref{Bogo}) results from Eq. (\ref{ccc}) since $c'$ is a positive unit vector, $ \langle c'\cdot c'\rangle=1$  \end{demo} 
A similar result holds in R.II for the equivalent $c$-bases (with the same vectors $c$ and $c'$). Therefore,   Eq. (\ref{Bogo}) guarantees that:
\begin{cor}
The Bogolyubov transformations conserve the canonical commutation relations (\ref{comab}) and (\ref{comabE}) .
\end{cor}
On the contrary, the vacuum states are strongly dependent on the choice of the unit vector vectors $c$. Thus, each pair of equivalent $c$-bases of  R.I and R.II  defines its own vacuum state, $|c\rangle=|(c_1,c_2)\rangle$, which satisfies $a_c|c\rangle=b_c|c\rangle=0$. 

On the other hand, Eq. (\ref{Bogo}) shows that the transition coefficients of any non-trivial Bogolyubov transformation have Planckian forms that allow one to define the (relative) Hawking temperature 
\begin{equation}
T= \frac{\omega\nu}{2}\left(\ln \left|\frac{ \langle c'\cdot c\rangle}{\langle c'\cdot\bar{ c}\rangle}\right|\right)^{-1}\,,
\end{equation} 
interpreted as the temperature of the thermal bath of particles (of ground energy $\omega\nu$)  measured by an observer ($c$)  in the vacuum state prepared by his partner ($c'$).   

\subsection{PCN and PCT invariant bases for $\mu>\lambda$} \label{4_2}
\label{subs}
Let us consider the particular base in R.I whose  mode functions,  normalized in the momentum scale according to Eq. (\ref{JuJu}) read 
\begin{equation}\label{modeP}
u_{\nu,{\bf p}}(t,{\bf x})=\sqrt{\frac{\pi}{\omega}}\frac{1}{\sqrt{2\sinh\pi\nu}}\frac{(-\omega t)^{\frac{3}{2}}}{(2\pi)^{\frac{3}{2}}} J_{i\nu}(-p t) e^{i\bf{x}\cdot{\bf p}} \in {\cal F}_+\,,
\end{equation} 
where $p=|{\bf p}|$ and $J$ are  Bessel functions \cite{GR}. The equivalent mode functions of R.II may be derived as in Ref. \cite{cota_crucean_pop},
\begin{equation}\label{modeE}
u_{\nu,E,{\bf n}}(t, {\bf x})=\sqrt{\frac{\omega}{2}}\frac{1}{\sqrt{2\sinh\pi\nu}}\frac{(-\omega t)^{\frac{3}{2}}}{(2\pi)^{\frac{3}{2}}} \int_{0}^{\infty} ds\, s^{\frac{1}{2}-i\frac{E}{\omega}} \, J_{i\nu} (-s t)\,
 e^{i \omega s {\bf n}\cdot{\bf x}}\in {\cal F}_+\,.
\end{equation}
\begin{theor}\label{PCTN}
The mode functions $u_{\nu,{\bf p}}$ and $u_{\nu,E,{\bf n}}$ transform under elementary discrete transformations as
 \begin{equation}\label{transu}
 \begin{array}{lll}
 P u_{\nu,{\bf p}}=u_{\nu,-{\bf p}}&\quad &P u_{\nu,E,{\bf n}}=u_{\nu,E,-{\bf n}} \\
 T u_{\nu,{\bf p}}= e^{-\pi\nu+\frac{3}{2}i\pi}u_{\nu,{\bf p}}&\quad &T u_{\nu,E,{\bf n}}=e^{-\pi\nu+\frac{3}{2}i\pi} u_{\nu,E,{\bf n}} \\ 
 C u_{\nu,{\bf p}}=u^*_{\nu,{\bf p}}&\quad &C u_{\nu,E,{\bf n}}=u^*_{\nu,E,{\bf n}}\\
 N u_{\nu,{\bf p}}=u_{-\nu,{\bf p}}=-i u_{\nu,-{\bf p}}^*&\quad &N u_{\nu,E,{\bf n}}=u_{-\nu,E,{\bf n}}=-i u_{\nu,-E,-{\bf n}}^*\end{array}
\end{equation} 
\end{theor}
\begin{demo}
The first equations result straightforwardly from Eqs. (\ref{modeF}) and (\ref{modeE}). The second ones are obtained using  $J_{i\nu}(-x)=J_{i\nu}(x) e^{-\pi\nu}$ (for $x>0$), while the effect of  $N$ is due to the fact that   $J_{i\nu}^*(x)=J_{-i\nu}(x)$ for $x\in {\Bbb R}$
\end{demo} 
Notice that the last of  Eqs. (\ref{transu}) specify the action of  $N:{\cal F}_+\to {\cal F}_+$ by setting $\sqrt{-\sin\pi\nu}=i\sqrt{\sin\pi\nu}$. 

It is important to note that all the Bessel functions have a branch cut along the real negative semi-axis \cite{nist}. Therefore, in order to unambiguously define the action of the $T$ symmetry, it is necessary to specify which branch displays the desired effects of the antipodal transformation on the time-dependent part of the mode function. If $x>0$ and $\nu>0$, then the following prescription is implied, which selects different branches for the function
\begin{equation}
J_{ \pm i\nu}(-x)=\lim_{\epsilon \searrow 0} J_{\pm i\nu}(-x \pm i \epsilon)=J_{\pm i\nu}(-x \pm i0)\,,
\end{equation}
and generally if $f_{c\,\nu}(t,{\bf x})$ is a positive frequency mode function and $f^*_{c\,\nu}(t,{\bf x})$ the corresponding negative frequency one, then by the time reversal we mean
\begin{equation}
\begin{array}{c}
Tf_{c\,\nu}(t,{\bf x})= f_{c\,\nu}(t+i0,{\bf x}) \,,\\
Tf^*_{c\,\nu}(t,{\bf x})= f^*_{c\,\nu}(t-i0,{\bf x}) \,. 
\end{array}
\end{equation}

Starting with these properties we may look  for the discrete symmetries of the general linear combinations (\ref{modeF}) and   (\ref{modeFE}).  
\begin{theor}\label{PCTNf}
The mode functions $f_{c\,\nu,{\bf p}}$ and $f_{c\,\nu,E,{\bf n}}$ of the $c$-bases defined by $c=(c_1,c_2)$, transform under elementary discrete transformations as
 \begin{equation}\label{transf}
 \begin{array}{lll}
 P f_{c\,\nu,{\bf p}}=f_{c\,\nu,-{\bf p}}&\quad &P f_{c\,\nu,E,{\bf n}}=f_{c\,\nu,E,-{\bf n}} \\
 T f_{c\,\nu,{\bf p}}= e^{\frac{3}{2}i\pi}f_{c'\,\nu,{\bf p}}&\quad &T f_{c\,\nu,E,{\bf n}}=e^{\frac{3}{2}i\pi} f_{c'\,\nu,E,{\bf n}} \\ 
 C f_{c\,\nu,{\bf p}}=f^*_{c\,\nu,{\bf p}}=f_{\bar{c}\,\nu,-{\bf p}}&\quad &C f_{c\,\nu,E,{\bf n}}=f^*_{c\,\nu,E,{\bf n}}=f_{\bar{c}\,\nu,-E,-{\bf n}}\\
 N f_{c\,\nu,{\bf p}}=f_{c\,-\nu,{\bf p}}&\quad & N f_{c\,\nu, E,{\bf n}}=f_{c\,-\nu,E,{\bf n}}\\
~~~~~~~~~ =-if_{c''\,\nu,{\bf p}}&&~~~~~~~~~~~=-if_{c''\,\nu,E,{\bf n}}
 \end{array}
\end{equation} 
where $c'=(e^{-\pi\nu}c_1,e^{\pi\nu}c_2)$ and $c''=(-c_2,c_1)$ .
\end{theor}
\begin{demo} We use Eqs. (\ref{transu}) and the complex conjugated ones, observing that  $N u^*_{\nu,-{\bf p}}=i u_{\nu,{\bf p}}$ and 
$N u_{\nu,-E,-{\bf n}}^*=i u_{\nu,E,{\bf n}}$
\end{demo}

We  look first for elementary discrete symmetries observing  that there are no eigenfunctions of  $P$ and $C$.  The functions   $u_{\nu,{\bf p}}$ and $u_{\nu,E,{\bf n}}$  are  eigenfunctions of  $T$ corresponding to the degenerated eigenvalue  $ e^{-\pi\nu+\frac{3}{2}i\pi}$ which changes the norm of these functions. Consequently, we cannot speak about $P$, $C$ or $T$ invariance separately. On the other hand,  the $PC$ transformation which has the simple action $PC f_{c\,\nu,{\bf p}}= f_{\bar{c}\,\nu,{\bf p}}$  deals with  an expected  result which forbids  this symmetry. 
\begin{cor}\label{corPC}
The $PC$ invariant mode functions which satisfy  the eigenvalues problem $PCf_{c\,{\bf p}}=\eta_{PC}f_{c\,{\bf p}}$  correspond   to null vectors $c$ with $\langle c \cdot c\rangle=0$ for  $|\eta_{PC}|=1$.
\end{cor}
\begin{demo}
According to Theorem \ref{PCTN}, the eigenvalues problem has solutions only for $|\eta_{PC}|=1$ and $c_2=\eta_{PC}c^*_1$ but then  $|c_1|=|c_2|$ and  $\langle c \cdot c\rangle=0$ 
\end{demo}
Thus it is clear that there are no $PC$-invariant $c$-bases of R.I or R.II. 

It remains to study only  the  $PCN$ and $PCT$ symmetries. Let us start with the $PCN$ one.
\begin{cor}
The only  $PCN$-invariant $c$-bases of R.I and  R.II are  $\{u_{\nu,{\bf p}}\}$ and $\{u_{\nu,E,{\bf n}}\}$  whose mode functions obey  
\begin{equation}\label{PCNuu}
PCN u_{\bf p}=  i  u_{\bf p}\,,\quad PCN  u_{E,{\bf n}}=i  u_{-E,{\bf n}}\,.
\end{equation}
\end{cor}  
\begin{demo} The first of the above equations and its complex-conjugate,  $PCN u^*_{\bf p}=-  i  u^*_{\bf p}$, show that there are no other $PCN$-invariant linear combinations \end{demo}
This symmetry assures the closure conditions of the form  (\ref{complete})  for  both the $PCN$-bases under consideration here that define the $PCN$-vacuum. 

One remark that can be made is that the $PCN$ combined discrete symmetry selects a single set of mode functions, not a parametric family. Another observation is that this set of mode functions can be related to the ones used in Ref. \cite{sato}, or to the ones that designate the {\it in} modes of Ref. \cite{spradlin}. It is important to mention that in this last reference, these {\it in} modes are defined on the contracting Euclidean de Sitter patch, while we work on the expanding one. Therefore, the direct correspondent which gives a meaning to our $PCN$-invariant base in this case would be the {\it out} modes on the expanding Euclidean chart. The associated {\it out} vacuum corresponds to having no particles going to future null infinity (where proper time $t_{pr} \rightarrow \infty$ and thus conformal time $t \rightarrow 0$).

\begin{cor}
The eigenvalues problem  $PCT\,f_{c\,\nu,{\bf p}}=\eta f_{c\,\nu,{\bf p}}$ is solved as  
\begin{equation}\label{cPCT}
\eta=e^{-i(\theta_1+\theta_2+\frac{3}{2}\pi)}\to \quad c_1=\frac{e^{\frac{1}{2}\pi\nu+i\theta_1}}{\sqrt{2\sinh \pi\nu}}\,,\quad 
c_2= \frac{e^{-\frac{1}{2}\pi\nu+i\theta_2}}{\sqrt{2\sinh \pi\nu}}\,,
\end{equation}
where $c=(c_1,c_2)$ depends  on the arbitrary phases $\theta_1\,,\theta_2\in {\Bbb R}$. 
\end{cor}   
\begin{demo}
Using  Eqs. (\ref{transf}) we  derive  the system of equations  $c^*_2e^{\pi\nu}=e^{\frac{3}{2}i\pi}\eta c_1$ and $c^*_1 e^{-\pi\nu}=e^{\frac{3}{2}i\pi}\eta c_2$  giving   the solutions  (\ref{cPCT})  \end{demo}
We obtain thus two families  of  $PCT$-invariant $c$-bases in R.I and R.II, parametrised by  different  pairs of phases $(\theta_1,\theta_2)$. The mode functions of R.I read
\begin{eqnarray}\label{modeFtt}
f_{(\theta_1,\theta_2)\,\nu\,{\bf p}}(t,{\bf x})&=&\frac{1}{2}\,\sqrt{\frac{\pi}{\omega}}\frac{ (-\omega t)^{\frac{3}{2}}}{(2\pi)^{\frac{3}{2}}}e^{i\bf{x}\cdot{\bf p}}\nonumber\\
&&\times\left[\frac{e^{\frac{1}{2}\pi\nu+i\theta_1}}{\sinh \pi\nu}J_{i\nu}(-pt)+ \frac{e^{-\frac{1}{2}\pi\nu+i\theta_2}}{\sinh \pi\nu}J_{-i\nu}(-pt) \right]\,. 
\end{eqnarray} 
In general, different $PCT$-bases are not equivalent, since their Bogolyubov transformations are non-trivial for arbitrary phases.
\begin{theor}
Two $PCT$-bases having the phases $(\theta_1,\theta_2)$ and $(\theta'_1,\theta'_2)$ are equivalent only if 
$\theta'_2-\theta'_1=\theta_2-\theta_1$.
\end{theor}
\begin{demo}
Calculating the Bogolyubov coefficients with the constants (\ref{cPCT}) we obtain this result \end{demo} 

The mode functions (\ref{modeFtt}) satisfy $f_{(\theta_1,\theta_2)\,-\nu\,{\bf p}}=-f_{(\theta_2,\theta_1)\,\nu\,{\bf p}}$,  
which indicates that the $\nu$-symmetric ones must have  $\theta_1=0$ and $\theta_2=\pi$, taking the form
\begin{equation}\label{PCTRI}
f_{\bf p}(t,{\bf x})\equiv f_{(0,\pi)\,\nu\,{\bf p}}(t,{\bf x})=\frac{1}{2}\,\sqrt{\frac{\pi}{\omega}}\frac{ (-\omega t)^{\frac{3}{2}}}{(2\pi)^{\frac{3}{2}}} Z_{i\nu}(-pt)e^{i\bf{x}\cdot{\bf p}}\in {\cal F}_+\,,
\end{equation}
where  $Z_{i\nu}$ is  the $\nu$-symmetric Bessel function (\ref{Zfunction}). A similar result can be obtained for the equivalent basis of R.II, whose mode functions $f_{E,{\bf n}}$  are given in Ref. \cite{cota_crucean_pop}. 
\begin{rem}
The bases $\{f_{\bf p}\}$ and $\{f_{E,{\bf n}}\}$ are simultaneously $\nu$-symmetric and  $PCT$-invariant, 
\begin{equation}\label{PCTfp}
  PCT f_{\bf p}= i  f_{\bf p} \,, 
\quad  PCT f_{E,{\bf n}}= i  f_{-E,{\bf n}}\,.
\end{equation}
These correspond to the Euclidean or Bunch-Davies vacuum.
\end{rem}

The above equations guarantee the closure condition  of the $\nu$-symmetric $PCT$-bases  in the sense of definition \ref{complete}.

The relation between the complex constants $c_1,c_2$ and the MA parametrisation variables $\alpha,\beta,\gamma$ \cite{allen} is simply given by $c_1=e^{-i\gamma}\cosh \alpha$ and $c_2=-e^{i\gamma} e^{i\beta} \sinh \alpha$. Our condition of $\nu$-symmetry realizes the same result as the $\gamma=0$ assumption in the MA parametrisation. There, $e^{i\gamma}$ is just a phase factor and it can be discarded without affecting the physical result. 

Finally, it is worth pointing out that the Bogolyubov transformations between a $PCN$-basis (of R.I or R.II) and one of the $PCT$-bases, of any phases $(\theta_1,\theta_2)$,  have the coefficients  given by Eqs. (\ref{cPCT}). These are independent on ${\bf p}$ (or $E$ and ${\bf n}$) and  satisfy Eq. (\ref{Bogo})  having the Planckian form
\begin{equation}
|c_1|^2=\frac{e^{2\pi\nu}}{e^{2\pi\nu}-1}\,,\quad |c_2|^2=\frac{1}{e^{2\pi\nu}-1}\,,
\end{equation}
corresponding to the natural relative temperature $\frac{\omega}{2\pi}$ \cite{hawking}.
 
\subsection{$PCN$ and  $PCT$-invariant bases for $\mu<\lambda$} \label{sec43}

In this case we have another normalization rule given by Eq. (\ref{JuJu}) so that we must redefine the basic mode functions of R.I as
\begin{equation}\label{modeP1}
u_{\nu,{\bf p}}(t,{\bf x})=\sqrt{\frac{\pi}{\omega}}\frac{1}{2\sqrt{\sin\pi\nu}}\frac{(-\omega t)^{\frac{3}{2}}}{(2\pi)^{\frac{3}{2}}}\left[ J_{\nu}(-p t)-i J_{-\nu}(-p t)\right]\,e^{i\bf{x}\cdot{\bf p}}\,,
\end{equation} 
and similarly for the equivalent mode functions of R.II. These have now  the good orthogonality and completeness properties  (\ref{ortou1})-(\ref{comp2}), but their form leads to complicated calculations.  For this reason,  it is convenient to use a special parametrization, $c=c(\alpha,\vartheta)$ with $0<\vartheta<\pi$ and $\alpha\in {\Bbb R}$, defined as 
\begin{eqnarray}
c_1&=&\frac{1}{2\sqrt{\sin\vartheta}}\left(e^{\frac{1}{2}i\vartheta+\alpha}+i e^{-\frac{1}{2}i\vartheta-\alpha}\right)\,,\label{C1}\\
c_2&=&\frac{1}{2\sqrt{\sin\vartheta}}\left(-e^{\frac{1}{2}i\vartheta+\alpha}+i e^{-\frac{1}{2}i\vartheta-\alpha}\right)\,,
\label{C2}
\end{eqnarray}
 for all the general $c$-bases  (\ref{modeF}) and (\ref{modeFE}) we have to study here. With this parametrization the mode functions of the $c$-bases in R.I take the form
\begin{eqnarray}
f_{(\alpha,\vartheta)\,\nu,{\bf p}}&\equiv & f_{c\,\nu,{\bf p}}(t,{\bf x})=\sqrt{\frac{\pi}{\omega}}\frac{i}{2\sqrt{\sin\pi\nu\sin\vartheta}}\frac{(-\omega t)^{\frac{3}{2}}}{(2\pi)^{\frac{3}{2}}}\nonumber\\
&&\times \left[e^{-\frac{1}{2}i\vartheta-\alpha} J_{\nu}(-p t)-e^{\frac{1}{2}i\vartheta+\alpha} J_{-\nu}(-p t)\right]\,e^{i\bf{x}\cdot{\bf p}}\in {\cal F}_+\,,\label{modePteta}
\end{eqnarray}    
and similarly for the equivalent $c$-bases of R.II.  The mode functions of negative frequencies are  $f_{\bar{c}\,\nu,{\bf p}}=-if_{(\alpha,-\vartheta)\, \nu,{\bf p}}\in{\cal F}_-$, since  $\bar{c}(\alpha,\vartheta)=-ic(\alpha,-\vartheta)$. Thus we can work exclusively in this  parametrisation avoiding to use  the mode functions (\ref{modeP1}) which can be seen now as a particular case, $u_{\nu,{\bf p}}=e^{i\frac{\pi}{4}}f_{(0,\frac{\pi}{2})\,\nu,{\bf p}}$.  
\begin{theor}
For  two pairs of equivalent $c$-bases in R.I and R.II, of parameters $c=c(\alpha,\vartheta)$ and  $c'=c(\alpha',\vartheta')$,   Eqs. (\ref{scalarFF}) and  (\ref{scalarFFE})  hold with
\begin{equation}\label{ccaa}
\langle c'\cdot c\rangle=\frac{\cosh(\alpha'-\alpha)\sin\frac{1}{2}(\vartheta+\vartheta')+i \sinh(\alpha'-\alpha)\cos\frac{1}{2}(\vartheta+\vartheta')}{\sqrt{\sin\vartheta \sin\vartheta'}}\,.
\end{equation}
\end{theor}
\begin{demo}
Substituting the components  (\ref{C1}) and (\ref{C2}) in Eq. (\ref{cc}) we obtain the desired result  \end{demo}

The parametrization (\ref{modePteta}) is suitable for analysing the effects of the discrete symmetries.
\begin{theor}\label{PCTN1}
The mode functions $f_{(\alpha,\vartheta)\,\nu,{\bf p}}$ and $f_{(\alpha,\vartheta)\,\nu,E,{\bf n}}$ transform under elementary discrete transformations as
 \begin{equation}
 \begin{array}{lll}
 P f_{(\alpha,\vartheta)\,\nu,{\bf p}}=f_{(\alpha,\vartheta)\,\nu,-{\bf p}}&\quad &P f_{(\alpha,\vartheta)\,\nu,E,{\bf n}}=f_{(\alpha,\vartheta)\,\nu,E,-{\bf n}} \\
 T f_{(\alpha,\vartheta)\,\nu,{\bf p}}=&\quad &T f_{(\alpha,\vartheta)\,\nu,E,{\bf n}}=\\
~~ e^{\frac{3}{2}i\pi}\sqrt{\frac{\sin(\vartheta-2\pi\nu)}{\sin\vartheta}}f_{(\alpha,\vartheta-2\pi\nu)\,\nu,{\bf p}}&\quad &~~ e^{\frac{3}{2}i\pi}\sqrt{\frac{\sin(\vartheta-2\pi\nu)}{\sin\vartheta}} f_{(\alpha,\vartheta-2\pi\nu)\,\nu,E,{\bf n}} \\ 
 C f_{(\alpha,\vartheta)\,\nu,{\bf p}}=&\quad &C f_{(\alpha,\vartheta)\,\nu,E,{\bf n}}=\\
~~f^*_{(\alpha,\vartheta)\,\nu,{\bf p}}=i f_{(\alpha,-\vartheta)\,\nu,-{\bf p}}&\quad &~~f^*_{(\alpha,\vartheta)\,\nu,E,{\bf n}}=i f_{(\alpha,-\vartheta)\,\nu,-E,-{\bf n}}\\
 N f_{(\alpha,\vartheta)\,\nu,{\bf p}}=&\quad &N f_{(\alpha,\vartheta)\,\nu,E,{\bf n}}=\\
~~f_{(\alpha,\vartheta)\,-\nu,{\bf p}}=-f_{(-\alpha,-\vartheta)\,\nu,{\bf p}}&\quad &~~f_{(\alpha,\vartheta)\,-\nu,E,{\bf n}}=- f_{(-\alpha,-\vartheta)\,\nu,E,{\bf n}}
 \end{array}
\end{equation} 
\end{theor}
\begin{demo}
Now the Bessel functions have a  real-valued index so that $J_{\nu}^*(x)=J_{\nu}(x)$.  The second equations are obtained taking into account that  $J_{\nu}(-x)=J_{\nu}(x) e^{i\pi\nu}$, while the other ones result directly from Eq. (\ref{modePteta}) \end{demo}
\begin{cor}
The $c$-bases   $\{f_{c\,\nu,{\bf p}}\}$ and $\{f_{c\,\nu,E,{\bf n}}\}$  are $PCN$-invariant only for $c=c(0,\vartheta)$, with $\alpha=0$ and any $\vartheta\in(0,\pi)$, 
\begin{equation}\label{PCNuu2}
PCN  f_{(0,\vartheta)\,\nu,{\bf p}}= - i   f_{(0,\vartheta)\,\nu,{\bf p}}\,,\quad PCN   f_{(0,\vartheta)\,\nu,E,{\bf n}}=-i f_{(0,\vartheta)\,\nu,-E,{\bf n}}\,.
\end{equation}
\end{cor}  
Hereby we  see that in the domain $\mu<\lambda$ we can use any $PCN$-basis with $c=c(0,\vartheta)$  for determining  specific vacuum states.  Two arbitrary  bases of this type with  $c=c(0,\vartheta)$ and  $c'=c(0,\vartheta')$  are related between themselves through the Bogolyubov transformation whose coefficients 
\begin{equation}
\langle c'\cdot c\rangle=\frac{\sin\frac{1}{2}(\vartheta'+\vartheta)}{\sqrt{\sin\vartheta\sin\vartheta'}}\,,\quad  \langle c'\cdot \bar{c} \rangle=-i\,\frac{\sin\frac{1}{2}(\vartheta'-\vartheta)}{\sqrt{\sin\vartheta\sin\vartheta'}}\,,
\end{equation} 
satisfy Eq. (\ref{Bogo}).

Furthermore,  we  look  for other discrete symmetries in  R.I (or R.II). We observe first that there are no $PC$-invariant $\vartheta$-bases since this invariance requires $\vartheta=0$ leading to a singularity.  It remains to study the  $PCT$ symmetry. 
\begin{cor}
The  $PCT$-invariant mode functions may have only $\vartheta=\pi\nu$, but any $\alpha\in{\Bbb R}$ 
\begin{equation}\label{PCTuu}
PCT  f_{(\alpha,\pi\nu)\,\nu,{\bf p}}= - i   f_{(\alpha,\pi\nu)\,\nu,{\bf p}}\,,\quad PCT   f_{(\alpha,\pi\nu)\,\nu,E,{\bf n}}=-i f_{(\alpha,\pi\nu)\,\nu,-E,{\bf n}}\,.
\end{equation}
\end{cor}   
\begin{demo}
Using  the results of Theorem \ref{PCTN1}, we find that $f$ and $PCT f$ may have the same angle only when $\vartheta=\pi\nu$ \end{demo}
We meet again a large class of $PCT$-bases which depend continuously on the parameter $\alpha$,  each one defining its own vacuum state. Given two arbitrary $PCT$-bases, with $c=c(\alpha,\pi\nu)$  and   $c'=c(\alpha',\pi\nu)$, we derive their Bogolyubov transformation having the coefficients
\begin{eqnarray}
\langle c'\cdot c\rangle &=&\cosh(\alpha'-\alpha)+i \sinh(\alpha'-\alpha)\,{\rm cot}(\pi\nu)\,,\\
\langle c'\cdot\bar{c}\rangle &=&-i \sinh(\alpha'-\alpha)\,{\rm csc}(\pi\nu)\,,
\end{eqnarray}
resulting from Eq.  (\ref{ccaa}). These satisfy the condition (\ref{Bogo}) since
\begin{equation}
|\langle c'\cdot c\rangle|^2=\frac{\cosh^2(\alpha'-\alpha)-\cos^2\pi\nu}{\sin^2\pi\nu}\,,\quad 
|\langle c'\cdot\bar{c}\rangle|^2=\frac{\sinh^2 (\alpha'-\alpha)}{\sin^2\pi\nu} \,.
\end{equation}
Other Bogolyubov transformations among $PCN$ and $PCT$ bases can also be calculated using Eq.  (\ref{ccaa}).
\begin{rem}
The $PCT$-invariant bases in R.I and R.II with $\alpha=0$ are just the standard ones corresponding to the Euclidean vacuum. These are simultaneously $PCT$-invariant and $\nu$-symmetric.
\end{rem}
The mode functions of these bases are denoted simply by $f_{\bf p}\equiv f_{(0,\pi\nu)\,\nu,{\bf p}}$ and $f_{E,{\bf n}}\equiv f_{(0,\pi\nu)\,\nu,E,{\bf n}}$, since these are similar to those of the case $\mu>\lambda$ being given by the same Eq. (\ref{PCTRI}), but with $Z_{\nu}$ instead of $Z_{i\nu}$.  
\begin{ex}{\bf 2: The massless case} When $\mu=0$ then $\nu=\lambda$ can be either  $\frac{1}{2}$ in the conformal coupling or   $\frac{3}{2}$ in the minimal one. In both these cases the Bogolyubov transformation between two $PCT$-bases (of parameters $\alpha$ and $\alpha'$)  has simple coefficients,  $ |\langle c'\cdot c\rangle|=\cosh(\alpha'-\alpha)$ and $ |\langle c'\cdot\bar{c}\rangle|=|\sinh(\alpha'-\alpha)|$
\end{ex}

\section{Spherical modes in R.III and R.IV} \label{r34}

They can be obtained by transforming the modes in R.I and R.II from Cartesian to spherical spatial coordinates, and by expanding the plane waves to corresponding spherical waves using Eq. (\ref{rayleigh}).

The eigenfunctions of the s.c.o. III corresponding to the eigenvalues  $\{\mu^2-\lambda^2, p,l,m\}$,  
\begin{eqnarray}
u_{\nu,p,l,m}(t,r,\theta,\phi)&=&\frac{1}{2}\sqrt{\frac{p\pi}{\omega}}\frac{e^{i m\frac{\pi}{2}}}{\sqrt{2\sinh\pi\nu}}\frac{(-\omega t)^{\frac{3}{2}}}{\sqrt{r}}\nonumber\\
&&\times  J_{i\nu}(-p t) J_{l+\frac{1}{2}}(p r) Y_{l,m}(\theta,\phi)\in {\cal F}_+ \,, \label{modeplm}
\end{eqnarray} 
are orthonormalized in the momentum scale
\begin{eqnarray}
\langle u_{\nu, p,l,m},u_{\nu,p',l',m'}\rangle=-\langle u^*_{\nu, p,l,m},u^*_{\nu,p',l',m'}\rangle & =& \delta(p-p')\delta_{ll'}\delta_{mm'}\,,\\\label{transw1A}
\langle u^*_{\nu, p,l,m},u_{\nu,p',l',m'}\rangle &=&0\,, \label{transw2A}
\end{eqnarray}
under the scalar product (\ref{SP2}). The equivalent eigenfunctions of s.c.o. III of eigenvalues $\{\mu^2-\lambda^2, E,l,m\}$,
\begin{eqnarray}
u_{\nu,E,l,m}(t,r,\theta,\phi)&=&\frac{1}{2\sqrt{2}} \frac{e^{i m\frac{\pi}{2}}}{\sqrt{2\sinh\pi\nu}} \frac{(-\omega t)^{\frac{3}{2}}}{\sqrt{r}}   Y_{l,m}(\theta,\phi)\nonumber\\ 
&&\times \int_0^\infty s^{-i\epsilon} J_{i\nu}(-\omega st) J_{l+\frac{1}{2}}(\omega rs) ds \in {\cal F}_+\,, \label{modeElm}
\end{eqnarray} 
are normalized in the energy scale,
\begin{eqnarray}
\langle u_{\nu, E,l,m},u_{\nu,E',l',m'}\rangle=-\langle u^*_{\nu, E,l,m},u^*_{\nu,E',l',m'}\rangle & =& \delta(E-E')\delta_{ll'}\delta_{mm'}\,,\\\label{transw1B}
\langle u^*_{\nu, E,l,m},u_{\nu,E',l',m'}\rangle &=&0\,. \label{transw2B}
\end{eqnarray}
While the modes of R.III can also be found just by solving the Klein-Gordon equation in the appropriate chart, those of R.IV require special considerations \cite{pascu}.

Being expressed in Euclidean coordinates, the temporal part of the modes is the same. Therefore the structure of the vacuum family is the same. The only thing that differs is the concrete action of the discrete symmetries on the mode functions.
For $\mu>\lambda$, this amounts to the following
\begin{theor}\label{PCTN2}
The mode functions $u_{\nu,p,l,m}$ and $u_{\nu,E,l,m}$ transform under elementary discrete transformations as
 \begin{equation}\label{transu_plm}
 \begin{array}{lll}
 P u_{\nu,p,l,m}=(-1)^l u_{\nu,p,l,m}&\quad &P u_{\nu,E,l,m}=(-1)^l u_{\nu,E,l,m} \\
 T u_{\nu,p,l,m}= e^{-\pi\nu+\frac{3}{2}i\pi}u_{\nu,p,l,m}&\quad &T u_{\nu,E,l,m}=e^{-\pi\nu+\frac{3}{2}i\pi} u_{\nu,E,l,m} \\ 
 C u_{\nu,p,l,m}=u^*_{\nu,p,l,m}&\quad &C u_{\nu,E,l,m}=u^*_{\nu,E,l,m}\\
 N u_{\nu,p,l,m}=u_{-\nu,p,l,m}&\quad &N u_{\nu,E,{\bf n}}=u_{-\nu,E,{\bf n}}\\
~~~~~~~~~~~ =-i u_{\nu,p,l,-m}^*&\quad &~~~~~~~~~~=-i  u_{\nu,-E,l,-m}^*
 \end{array}
\end{equation} 
\end{theor}
\begin{demo}
The difference from the previous cases is the action of the charge conjugation and parity on the spherical harmonics
\end{demo} 
Now we see that the bases $\{u_{\nu,p,l,m}\}$ and  $\{u_{\nu,E,l,m}\}$ play the role of $PCN$-invariant bases
\begin{equation}
PCN  u_{\nu,p,l,m}=i(-1)^l  u_{\nu,p,l,-m}\,, \quad  PCN  u_{\nu,E,l,m}=i(-1)^l  u_{\nu,-E,l,-m}\,,
\end{equation}  
but with eigenvalues depending on $l$.

The general eigenfunctions of s.c.o. III are of the form 
\begin{equation}
f_{(c_1,c_2)\,\nu,p,l,m}=c_1 u_{\nu,p,l,m}+c_2 u^*_{\nu,p,l,-m}\,,\quad c_1,\, c_2\in {\Bbb C}\,,
\end{equation} 
and for s.c.o IV we have
\begin{equation}
f_{(c_1,c_2)\,\nu,E,l,m}=c_1 u_{\nu,E,l,m}+c_2 u^*_{\nu,-E,l,-m}\,,\quad c_1,\, c_2\in {\Bbb C}\,,
\end{equation} 
\begin{theor}\label{PCTNf2}
The mode functions $f_{c\,\nu,p,l,m}$ and $f_{c\,\nu,E,l,m}$, of the $c$-bases defined by $c=(c_1,c_2)$, transform under elementary discrete transformations as
 \begin{equation}\label{transf2}
 \begin{array}{lll}
 P f_{c\,\nu,p,l,m}=(-1)^l f_{c\,\nu,p,l,m}&\quad &P f_{c\,\nu,E,l,m}=(-1)^l f_{c\,\nu,E,l,m} \\
 T f_{c\,\nu,p,l,m}= e^{\frac{3}{2}i\pi}f_{c'\,\nu,p,l,m}&\quad &T f_{c\,\nu,E,l,m}=e^{\frac{3}{2}i\pi} f_{c'\,\nu,E,l,m} \\ 
 C u_{c\,\nu,p,l,m}=f^*_{c\,\nu,p,l,-m}
 &\quad &C f_{c\,\nu,E,l,m}=f^*_{c\,\nu,E,l,m}\\
 N f_{c\,\nu,p,l,m}=f_{c\,-\nu,p,l,m}&\quad & N f_{c\,\nu, E,l,m}=f_{c\,-\nu,E,l,m}\\
~~~~~~~~~ =-i(-1)^m f_{c''\,\nu,p,l,m}&&~~~~~~~~~~~=-i (-1)^m f_{c''\,\nu,E,l,m}
 \end{array}
\end{equation} 
where $c'=(e^{-\pi\nu}c_1,e^{\pi\nu}c_2)$ and $c''=(-c_2,c_1)$ .
\end{theor}
\begin{demo} Straightforward by combining the elementary discrete transformations (\ref{transf2})
\end{demo}

Hereby we understand that for $\mu>\lambda$  the mode functions of R.III and R.IV have similar properties as those of R.I and R. II,  generating the same families of $c$-bases. Moreover, it is not difficult to verify that under threshold, when $\mu<\lambda$, we recover similar properties as in section \ref{subs}.  Therefore, we can conclude the following:
\begin{rem}
All the results concerning the different families of $c$-bases with given discrete symmetries obtained for R.I and R.II hold for R.III and R.IV too.
\end{rem}

A comment is in order here. The mode functions $u_{\nu,E,l,m}(t,r,\theta,\phi)$ coincide with those of the static chart (\ref{Ztst}) denoted by $u^S_{\nu,E,l,m}(t_s,r_s,\theta,\phi)$, in the region on which the latter are defined, via a simple coordinate transformation. In these conditions, the integral form in which $u_{\nu,E,l,m}$ was given resolves to a hypergeometric function:
 \begin{align}\label{static}
 u^S_{\nu,E,l,m}(t_s,r_s,\theta,\phi)\propto (1-\omega^2 r_s^2)^{\frac{iE}{\omega}} r_s^l e^{iEt_s} Y_{l,m}(\theta,\phi) \notag \\
\times {}_2 F_1\left( \frac{3}{4}+\frac{iE}{2\omega}+\frac{\nu}{2}+\frac{l}{2},\frac{3}{4}+\frac{iE}{2\omega}-\frac{\nu}{2}+\frac{l}{2};l+\frac{3}{2};\omega^2 r_s^2\right)\,.
 \end{align}
 However, the normalisation constant may  differ, since the scalar products expressed in the two charts are not equal, that of the static chart, 
\begin{equation} \label{SP3}
\langle f,f^\prime \rangle_s =i \int^{\frac{1}{\omega}}_0 \!\! \frac{dr_s}{1-\omega^2 r_s^2} \int_{\Omega_2}\sin \theta\,d\theta d\phi \, f^*(t_s,r_s,\theta,\phi ) \stackrel{\leftrightarrow}{\partial_{t_s}} f^\prime(t_s,r_s,\theta,\phi )
\,,
\end{equation}
being calculated along the hypersurface determined by $t_s=0$. 

\section{The unitary global representation V} \label{r5}

The solutions in this representation are best written using the global parametrisation (\ref{Ztau}) corresponding to a foliation with spatial surfaces of positive constant curvature, where the metric is
\begin{equation}
ds^2=d\tau^2- \cosh^2  \tau\, [d\chi^2- \sin^2 \chi(d\theta^2 - \sin^2 \theta d\phi^2)]\,.
\end{equation}
In this chart, the variables of the Klein-Gordon equation  can be separated  obtaining a temporal Legendre equation which can be solved in terms of associate Legendre functions of real argument in the domain $(-1,1)$ \cite{GR}, $P^{\pm i\nu}_{\Lambda+\frac{1}{2}}(\pm \tanh  \tau)$, $Q^{\pm i\nu}_{\Lambda+\frac{1}{2}}(\pm \tanh  \tau)$, referred sometimes as Ferrers functions \cite{nist}. There are many ways to write down the solutions as  linear combinations of these functions  leading to different definitions of  the vacuum. For instance, the one of $P^{\pm i\nu}_{\Lambda+\frac{1}{2}}$ leads to the solutions of Mottola \cite{mottola}, while a combination of $P$ and $Q$ was used in Ref. \cite{allen} (for the massless case), and subsequently by Allen and Folacci \cite{folacci} for $\mu<\lambda$. We note that the selection of  the Euclidean vacuum in Ref. \cite{folacci} is based on the properties of the two-point functions which are  assumed  to be $O(1,4)$-invariant and of Hadamard form. However, in what follows we study exclusively the discrete symmetries of different linear combinations of mode functions of R.V.

\subsection{$PCT$-bases for $\mu>\lambda$}

For $\mu>\lambda$ we consider the  eigenfunctions of  positive frequencies  of the s.c.o. V, corresponding to the eigenvalues $\{\mu^2-\lambda^2, \Lambda(\Lambda+2),l(l+1),m\}$,  
\begin{eqnarray}\label{u_glob}
u_{\nu,\Lambda,l,m}\!\!\!&(\tau,\chi,\theta,\phi) =e^{i m\frac{\pi}{2}}\sqrt{\frac{\pi
} {2\sinh \pi \nu}} \cosh^{-3/2} \tau \nonumber\\ 
& \times P^{-i\nu}_{\Lambda+\frac{1}{2}}(\tanh  \tau) Y_{\Lambda,l,m}(\chi,\theta,\phi)\,,
\end{eqnarray}
 orthonormalized under a specific scalar product on the Cauchy surface $\tau={\rm const.}$
\begin{equation} \label{SP4}
\langle f,f^\prime \rangle_{\tau} =i \int_{\Omega_3} \!\! \cosh^3 \tau\sin^2\chi\sin\theta\,d\chi d\theta d\phi \, f^*(\tau,\chi,\theta,\phi ) \stackrel{\leftrightarrow}{\partial_{\tau}} f^\prime(\tau,\chi,\theta,\phi )
\,.
\end{equation}
\begin{theor}\label{PCTN3}
The mode functions $u_{\nu,\Lambda,l,m}$ transform under elementary discrete transformations as
 \begin{equation}\label{transu_Llm}
 \begin{array}{l}
 P u_{\nu,\Lambda,l,m}=(-1)^l u_{\nu,\Lambda,l,m} \\
 T u_{\nu,\Lambda,l,m}=i (-1)^l( A_{\nu} u_{\nu,\Lambda,l,m} + B_{\nu} u_{-\nu,\Lambda,l,m})\\ 
 C u_{\nu,\Lambda,l,m}=u^*_{\nu,\Lambda,l,m} \\
 N u_{\nu,\Lambda,l,m}=u_{-\nu,\Lambda,l,m}=-i u_{\nu,\Lambda,l,-m}^*\end{array}
\end{equation} 
where $A_{\nu}$ and $B_{\nu}$ are given by
 \begin{equation}\label{AB}
 \begin{array}{l}
 A_{\nu}=-A_{-\nu}=-i{\rm csch} \pi \nu\,, \\
 B_{\nu}=-B^*_{-\nu}=i e^{-i\Phi_{\nu}} \coth \pi \nu \,,\\
 \Phi_{\nu}=-\Phi_{-\nu}= \arg \frac{\Gamma(\Lambda+ 3/2+i \nu)}{\Gamma(\Lambda+ 3/2 -i \nu)}\,.
 \end{array}
 \end{equation} 
\end{theor}
\begin{demo}
The action of $C$, $P$, $T$ on the hyperspherical harmonics and relations between the Legendre functions (see the Appendix B)
\end{demo} 

Hereby we deduce that the set of functions (\ref{u_glob}) satisfy
\begin{equation}
PCN  u_{\nu,\Lambda,l,m}=i (-1)^l  u_{\nu,\Lambda,l,-m}\,,
\end{equation}
which means that they play the role of  a $PCN$-basis even though the above  equation does not satisfy exactly Def. \ref{d4}, as long as its eigenvalue depends on $l$.  

A simple transformation of the associated Legendre function from our $PCN$-basis to the hypergeometric function shows the precise equivalence between these $PCN$-basis modes and the {\it out} modes of Mottola \cite{mottola}. As in the case of section \ref{4_2}, again we find that also in this case the $PCN$-basis has the meaning of a mode function set corresponding to the {\it out} vacuum.

The general eigenfunctions of s.c.o. V are of the form 
\begin{equation}\label{modeF3}
f_{(c_1,c_2)\,\nu,\Lambda,l,m}=c_1 u_{\nu,\Lambda,l,m}+c_2 u^*_{\nu,\Lambda,l,-m}=c_1 u_{\nu,\Lambda,l,m}+i c_2 u_{-\nu,\Lambda,l,m}\,,
\end{equation}
where $c_1,\, c_2\in {\Bbb C}$. 
\begin{theor}\label{PCTNf3}
The mode functions $f_{c\,\nu,\Lambda,l,m}$ and of the $c$-bases defined by $c=(c_1,c_2)$, transform under elementary discrete transformations as
 \begin{equation}\label{transf3}
 \begin{array}{l}
 P f_{c\,\nu,\Lambda,l,m}=(-1)^l f_{c\,\nu,\Lambda,l,m} \\
 T f_{c\,\nu,\Lambda,l,m}=i (-1)^l f_{c'\,\nu,\Lambda,l,m} \\ 
 C f_{c\,\nu,\Lambda,l,m}=f^*_{c\,\nu,\Lambda,l,m} \\
 N f_{c\,\nu,\Lambda,l,m}=f_{c\,-\nu,\Lambda,l,m}= -i f_{c''\,\nu,\Lambda,l,m}
 \end{array}
\end{equation} 
where $c'=(A_{\nu} c_1-i  B_{\nu}^* c_2 ,i B_{\nu} c_1+ A_{\nu} c_2)$ and $c''=(-c_2,c_1)$ .
\end{theor}
\begin{demo} Straightforward, by combining the elementary discrete transformations (\ref{transf3})
\end{demo}
\begin{cor}
The eigenvalues problem $PCT f_{c\,\nu,\Lambda,l,m}=\eta f_{c\,\nu,\Lambda,l,m}$ is solved  by 
$\eta=e^{i \varphi}$ and $c=(c_1,c_2)$ with
\begin{eqnarray}
c_1&=&-\frac{1}{\sqrt{1 +\sin (\varphi-\Phi_{\nu}) \sinh 2\pi\nu- \cosh 2\pi \nu}}\,,\\
c_2&=& \frac{-e^{-i\Phi_{\nu}}\cosh \pi \nu+i e^{-i\varphi}\sinh\pi\nu}{\sqrt{1+ \sin (\varphi-\Phi_{\nu}) \sinh 2\pi\nu- \cosh 2\pi \nu}}\,,
\end{eqnarray}
for any $\varphi\in {\Bbb R}$.
\end{cor}
\begin{demo} This problem is equivalent to the system
\begin{equation}
-i \eta c_1=i  c_1^* B^*-c_2^* A\,, \quad
-i \eta c_2= c_1^* A+i c_2^*\,,
\end{equation}
which can be solved for any $|\eta|=1$ \end{demo}  

We obtain thus a one-parameter family of $PCT$-bases among them we can select that corresponding to the Euclidean vacuum.
\begin{rem}
 For $\varphi=\Phi_{\nu}+\frac{\pi}{2}$ the $PCT$-basis $\{ f_{\hat c\,\nu,\Lambda,l,m}\}$ with
\begin{equation}
\hat c_1=\frac{e^{\frac{\pi \nu}{2}}}{\sqrt{2\sinh \pi \nu}}\,, \quad
\hat c_2=-\frac{e^{-i\Phi_{\nu}}e^{-\frac{\pi \nu}{2}} }{\sqrt{2\sinh \pi \nu}}\,,
\end{equation}
 is $\nu$-symmetric,  $ f_{ \hat c\,-\nu,\Lambda,l,m}= e^{i\Phi_{\nu}}  f_{ \hat c\,\nu,\Lambda,l,m}$, 
corresponding to the Euclidean vacuum since its (above)  Bogolyubov coefficients  with respect to  the $PCN$-basis (\ref{u_glob})  correspond to the relative temperature $\frac{\omega}{2\pi}$.
\end{rem}
Thus we find the same mechanism of selecting the Euclidean vacuum as before.
This solution can be rewritten in the more elegant form
\begin{eqnarray}
f_{\hat c\,\nu,\Lambda,l,m}\!\!\!&(\tau,\chi,\theta,\phi)
=e^{i m\frac{\pi}{2}} \cosh^{-3/2} \tau\, 
\frac{\sqrt{\pi}}{2} e^{-\frac{\pi\nu}{2}} \, Y_{\Lambda,l,m}(\chi,\theta,\phi)\nonumber\\
&\times \left[ P^{-i\nu}_{\Lambda+\frac{1}{2}}(\tanh \tau)- \frac{2i}{\pi} Q^{-i\nu}_{\Lambda+\frac{1}{2}}(\tanh \tau)
\right] \,. \label{sol_pct}
\end{eqnarray}

Finally, we note that our $PCT$ -invariant  Euclidean vacuum  is  equivalent with the $CT$-symmetric ones of   Refs. \cite{allen,bousso} since in R.V  the role of parity is  somewhat hidden by the spherical symmetry so that this affects only phase factors.

\subsection{$PCT$-bases for $\mu<\lambda$}

In this case we follow the same procedure as in subsection \ref{sec43}, but starting directly with the general linear combinations 
\begin{eqnarray}
f_{c\,\nu,\Lambda,l,m}\!\!\!&(\tau,\chi,\theta,\phi)
=e^{i m\frac{\pi}{2}} \cosh^{-3/2} \tau\, Y_{\Lambda,l,m}(\chi,\theta,\phi)\nonumber\\
&\times \left[c_1 P^{\nu}_{\Lambda+\frac{1}{2}}(\tanh  \tau)+c_2 P^{-\nu}_{\Lambda+\frac{1}{2}}(\tanh  \tau)\right] \,, \label{fluV}
\end{eqnarray}
with $c=(c_1$, $c_2)\in{\Bbb C}^2$. Then the normalization condition 
\begin{equation}
\langle f_{c\,\nu,\Lambda,l,m}, f_{c\,\nu,\Lambda',l',m'}\rangle_{\tau}=\delta_{\Lambda,\Lambda'}\delta_{l,l'}\delta_{m,m'}\,,
\end{equation}
with respect to the scalar product (\ref{SP4}) requires 
\begin{equation}\label{normRV}
c_1^*c_2-c_2^*c_1=\frac{i\pi}{2\sin\pi\nu}\,,
\end{equation}
as it results from Eq. (\ref{WPP}). 
\begin{theor} The eigenvalues problem $PCTf_{c\,\nu,\Lambda,l,m}=\eta\, f_{c\,\nu,\Lambda,l,m}$ is solved for any $\eta=e^{i\varphi}$ and
\begin{eqnarray}
c_1&=&\frac{\sqrt{\pi}}{2}\,e^{-\frac{1}{2}\beta_{\nu}}\frac{\sqrt{\cos\pi\nu}}{\sin\pi\nu\sqrt{\sin\varphi}}\,,\\
c_2&=&-\frac{\sqrt{\pi}}{2}\,e^{\frac{1}{2}\beta_{\nu}}\frac{1+e^{-i\varphi}\sin\pi\nu}{\sin\pi\nu\sqrt{\cos\pi\nu}\sqrt{\sin\varphi}}\,,
\end{eqnarray}
where 
\begin{equation}
\beta_{\nu}=-\beta_{-\nu}=\ln\left[\frac{\Gamma(\Lambda+\frac{3}{2}+\nu)}{\Gamma(\Lambda+\frac{3}{2}-\nu)}\right]\,.
\end{equation}
\end{theor}
\begin{demo}
The transformation $PT$ changes $\tau\to -\tau$ and $\cos\chi\to -\cos\chi$. Then, according to Eqs. (\ref{Yglob}), (\ref{Ghege}) and (\ref{P_x})  we find the equivalent system
\begin{eqnarray}
c_1^*{\rm csc\,}\pi\nu+c_2^*e^{-\beta_{\nu}}\cot\pi\nu&=&-\eta c_1\,,\\
 c_1^*e^{\beta_{\nu}}\cot\pi\nu+c_2^*{\rm csc\,}\pi\nu&=&\eta c_2\,,
\end{eqnarray}
which is solved using the normalization condition (\ref{normRV})
\end{demo}
Results a family of $PCT$-bases depending on the angular parameter $\varphi$.
\begin{cor}
For $\varphi=\pi(\nu+\frac{1}{2})$  the $PCT$-invariant mode functions, $f_{ \hat c\,\nu,\Lambda,l,m}$, with
\begin{equation} \label{chat}
\hat c_1=\frac{\sqrt{\pi}}{2}\frac{e^{-\frac{1}{2}\beta_{\nu}}}{\sin\pi\nu}\,,\quad \hat c_2=-\frac{\sqrt{\pi}}{2}\frac{e^{\frac{1}{2}\beta_{\nu}}e^{-i\pi\nu}}{\sin\pi\nu}\,,
\end{equation}
 are, in addition, $\nu$-symmetric since $f_{\hat c\,-\nu,\Lambda,l,m}=e^{i\pi\nu} f_{ \hat c\,\nu,\Lambda,l,m}$.
\end{cor}
Therefore, as in the previous cases, we can assume that the  $PCT$-basis $\{ f_{ \hat c\,\nu,\Lambda,l,m}\}$ corresponds of the Euclidean vacuum. These mode functions  can also be expressed in the form
\begin{eqnarray}
f_{\hat c\,\nu,\Lambda,l,m}\!\!\!&(\tau,\chi,\theta,\phi)
=i e^{i m\frac{\pi}{2}} \cosh^{-3/2} \tau\, 
\frac{\sqrt{\pi}}{2}e^{\frac{\beta_\nu}{2}} Y_{\Lambda,l,m}(\chi,\theta,\phi) \, \nonumber\\
&\times \left[ P^{-\nu}_{\Lambda+\frac{1}{2}}(\tanh \tau)- \frac{2i}{\pi} Q^{-\nu}_{\Lambda+\frac{1}{2}}(\tanh \tau)
\right] \,, \label{sol_pct2}
\end{eqnarray}
which, excepting some normalisation factors and phases, is similar to that of the solutions valid over the threshold (\ref{sol_pct}). It is also in agreement with the solutions corresponding to the Euclidean vacuum in the literature  \cite{tagirov,garriga}.

\begin{ex}{\bf  3: Massless conformally coupled particle} This case is of interest, being frequently discussed in the literature \cite{allen,folacci} because of some technical advantages of the particular value  $\nu=\frac{1}{2}$.  A suitable parametrization of the normalized mode functions of  ${\cal F}_+$ is now 
 \begin{eqnarray}
& f_{(\alpha,\vartheta)\,\frac{1}{2},\Lambda,l,m}(\tau,\chi,\theta,\phi) =\frac{e^{i m\frac{\pi}{2}}}{2}\sqrt{\frac{\pi
} {2 \sin \vartheta}} \cosh^{-3/2} \tau\, Y_{\Lambda,l,m}(\chi,\vartheta,\phi)\nonumber\\
&\times \left[e^{-\frac{1}{2}i\vartheta-\alpha} P^{\frac{1}{2}}_{\Lambda+\frac{1}{2}}(\tanh  \tau)-e^{\frac{1}{2}i\vartheta+\alpha} P^{-\frac{1}{2}}_{\Lambda+\frac{1}{2}}(\tanh  \tau)\right]\,. 
\end{eqnarray}
\begin{theor}\label{PCTNg}
The mode functions $f_{(\alpha,\vartheta)\,\frac{1}{2},\Lambda,l,m}$ transform under elementary discrete transformations as
 \begin{equation}
 \begin{array}{lll}
 P f_{(\alpha,\vartheta)\,\frac{1}{2},\Lambda,l,m}=(-1)^lf_{(\alpha,\vartheta)\,\frac{1}{2},\Lambda,l,m}& \\
 T f_{(\alpha,\vartheta)\,\frac{1}{2},\Lambda,l,m}= i(-1)^{l+1}f_{(\alpha,\vartheta)\,\frac{1}{2},\Lambda,l,m}& \\ 
 C f_{(\alpha,\vartheta)\,\frac{1}{2},\Lambda,l,m}=f^*_{(\alpha,\vartheta)\,\frac{1}{2},\Lambda,l,m}=i f_{(\alpha,-\vartheta)\,\frac{1}{2},\Lambda,l,-m}\\
 N f_{(\alpha,\vartheta)\,\frac{1}{2},\Lambda,l,m}= f_{(\alpha,\vartheta)\,-\frac{1}{2},\Lambda,l,m}=-f_{(-\alpha,-\vartheta)\,\frac{1}{2},\Lambda,l,m}
 \end{array}
\end{equation}
\end{theor} 
\begin{demo}
The argument reflection of Legendre functions of half-integer order leads to just one term in (\ref{P_x}), so the treatment of this case is more similar to the one in subsection \ref{subs} \end{demo} 
Therefore,  there is one extra parameter $\alpha$ that can define the $PCT$-invariant family.  In  this case, the $PCN$-invariant combinations have $\alpha=0$, while the $PCT$ ones must have $\vartheta=\pi/2$ and $\alpha=\beta_\nu/2$, in order to become a particular case (up to a phase factor) of the modes given by the coefficients (\ref{chat})
\end{ex} 

\section{Conclusion}

The quantum modes on a curved background are of primal importance to a field theory. In the de Sitter manifold case, which conveniently has maximal symmetry- one can obtain these modes as eigenfunctions of a system of commutative operators with physical meaning, that also have the function of separating the Klein-Gordon equation. However, the quantum modes are not determined fully by these operators. For the case of de Sitter spacetime, we propose a criterion to resolve this ambiguity: we have shown that invariance with regard to a set of discrete symmetries is sufficient to supplement the incompleteness of the commutative set of operators and thus select a certain vacuum. Specifically, as discrete symmetries, we investigated invariance with respect to combinations of parity ($P$), charge conjugation ($C$), antipodal transformation ($T$), and an additional symmetry ($N$)- which is essentially a reflection of mass parameter $\nu$ for a specific choice of the modes.
For scalar fields over a certain mass threshold , there are two distinct bases that stand out- the $PCT$-invariant and $\nu$-symmetric one, corresponding to the Euclidean vacuum, and the $PCN$- invariant basis, which we argue that have the meaning of modes corresponding to so-called {\it out} vacua from the literature. The thermality relation between the states defined by the two bases is the well known one. \cite{hawking}

On the other hand, for fields under the mass threshold, there is not only one $PCN$-invariant set of mode functions, but a family of them. We proposed case by case parametrisations which exhibit the multitude of possibilities for obtaining various Bogolyubov coefficients between the aforementioned invariant bases.

The results are in agreement to earlier ones \cite{allen,schomblond,mottola} featuring the vacuum family, but with the tool of discrete symmetries as only selection criterion. In fixing the Euclidean vacuum, this leads to same result as imposing a condition for the Hadamard function requiring it to have same singularity stricture for neighbouring points as in the Minkowski setting. In our opinion, the former criterion is more convenient, due to its straghtforward character.

Finally, we must specify that our principal goal was the study of the structure of different families of mode functions ("$c$-bases"), induced by combinations of discrete symmetries, rather than the general features of Bogolyubov transformations between members of different such families ("representations"). This aspect also may be of interest \cite{mishima,nakayama,singh} and could be discussed elsewhere.

\appendix

\subsection*{Appendix A:  Spherical coordinates}

\setcounter{equation}{0} \renewcommand{\theequation}
{A.\arabic{equation}}

Here we consider two global spherical charts. The first one is the conformally flat chart  having the  coordinates $\{t, r,\theta,\phi\}$ associated to  the Cartesian ones $\{t, {\bf x}\}$,  defined by Eq. (\ref{Zx}) for any $t\in {\Bbb R}$.   Another global chart has the coordinates $\{\tau,\chi,\theta,\phi\}$ which satisfy
\begin{eqnarray} 
\omega z^0&=&\sinh\tau\,,\nonumber\\
\omega z^1&=&\cosh\tau\sin\chi\sin\theta\cos\phi\,, \nonumber\\
\omega z^2&=&\cosh\tau\sin\chi\sin\theta\sin\phi\,, \label{Ztau}\\
\omega z^3&=&\cosh\tau\sin\chi\cos\theta\,, \nonumber\\
\omega z^4&=&\cosh\tau\cos\chi\,.\nonumber
\end{eqnarray}
Moreover, there are  two static charts of coordinates $\{t_s,r_s,\theta,\phi\}_{\sigma}$ (with $\omega|r_s|< 1$) and $\sigma=\pm1$ such that
\begin{eqnarray}
\omega z^0&=&\sqrt{1-\omega^2 r_s^2}\sinh t_s\,, \nonumber\\
z^1&=&r_s \sin\theta\cos\phi\,, \nonumber\\
z^2&=&r_s \sin\theta\sin\phi\,, \label{Ztst}\\
z^3&=&r_s \cos\theta\,, \nonumber\\
\omega z^4&=&\sigma \sqrt{1-\omega^2 r_s^2}\cosh t_s\,.\nonumber
\end{eqnarray}
The static chart with $\sigma=1$ covers the domain $z^4>0$, while that of $\sigma=-1$ the opposite one with $z^4<0$.

The parity acts similarly in these charts transforming $\theta \to \pi-\theta$ and $\phi\to \phi+\pi$.   The $PT$ transformation (changing the signs of $z^0$ and $z^4$) has to transform $\tau\to -\tau$ and $\chi\to \pi-\chi$ in the global chart and 
$t_s\to -t_s$ and $\sigma\to -\sigma$ for the static ones.

\subsection*{Appendix B: Some properties of special functions}

\setcounter{equation}{0} \renewcommand{\theequation}
{B.\arabic{equation}}

\subsubsection*{Bessel functions}
The Bessel functions \cite{GR} of any  indices and real-valued  arguments $J_{\nu}(s)\,, s\in{\Bbb R}$ , have the non-vanishing Wronskians 
\begin{equation}\label{JuJu}
W\left[J_{-\nu},J_{\nu}\right](s)=J_{-\nu}(s) \stackrel{\leftrightarrow}{\partial_{s}}J_{\nu}(s)=\frac{2\sin\pi\nu}{\pi s}\,. 
\end{equation} 
When we use  the Hankel functions  $H^{(1,2)}_{\nu}(s)$, it is convenient to  define the new functions  
\begin{equation}\label{Zfunction}
Z_{\nu}(s)=e^{\frac{1}{2}i\pi \nu}H^{(1)}_{\nu}(s)\,,
\end{equation}
since these  are  $\nu$-symmetric, $Z_{-\nu}=Z_{\nu}$, and for imaginary indices satisfy 
\begin{equation}\label{ZuZu}
Z_{i\nu}^*(s) \stackrel{\leftrightarrow}{\partial_{s}}Z_{i\nu}(s)=W[H_{i\nu}^{(2)},H_{i\nu}^{(1)}](s)=\frac{4i}{\pi
s}\,,\quad \nu\in{\Bbb R}\,.
\end{equation}

\subsubsection*{ $S^2$ and $S^3$  harmonics}

The expansion of plane waves into spherical waves reads
\begin{equation}\label{rayleigh}
e^{i\vec{q}\cdot\vec{x}}=(2\pi)^{\frac{3}{2}} \frac{1}{\sqrt{qr}}\sum\limits_{l=0}^\infty \sum\limits_{m=-l}^l i^l J_{l+\frac{1}{2}}(qr) Y_{l m}(\theta,\phi) Y^*_{l m}(\theta_{\vec{q}},\phi_{\vec{q}})\,.
\end{equation}
The hyperspherical harmonics are
\begin{equation}\label{Yglob}
Y_{\Lambda,l,m}(\chi,\theta,\phi)= l! 2^l\sqrt{\frac{2(\Lambda+1)(\Lambda-l)!}{\pi(\Lambda+l-1)}} (\sin \chi)^{l} C_{\Lambda-l}^{l-1} (\cos \chi) Y_{l,m} (\theta,\phi)\,,
\end{equation}
where $C^\alpha_n(x)$ are Gegenbauer orthogonal polynomials that have the following reflection property:
\begin{equation}\label{Ghege}
C^{\alpha}_n(-x)=(-1)^n C^{\alpha}_n(x)\,.
\end{equation}

\subsubsection*{ Legendre functions}

In the domain $|x|<1$ we use the associate Legendre functions \cite{GR,nist} of the first kind  $P^{\pm\nu}_{\Lambda+\frac{1}{2}}(x)$, with integer $\Lambda$ and any $\nu$,  since these form a complete system satisfying the Wronskian condition
\begin{equation}\label{WPP}
W\left[P^{-\nu}_{\Lambda+\frac{1}{2}},P^{\nu}_{\Lambda+\frac{1}{2}} \right](x)=\frac{2 \sin \nu \pi}{\pi} \frac{1}{1-x^2}\,. 
\end{equation} 
Then  the  functions of second kind read 
\begin{equation}
Q^{\nu}_{\Lambda+\frac{1}{2}}(x)=\frac{\pi}{2\sin\pi\nu}\left[\cos\pi\nu\,P^{\nu}_{\Lambda+\frac{1}{2}}(x)-
\frac{\Gamma(\Lambda+\frac{3}{2}+\nu)}{\Gamma(\Lambda+\frac{3}{2}-\nu)}P^{-\nu}_{\Lambda+\frac{1}{2}}(x)\right]\,,
\end{equation}
while the reflection of argument gives
\begin{equation}\label{P_x}
P^{\nu}_{\Lambda+\frac{1}{2}}(-x)=\frac{(-1)^{\Lambda}}{\sin\pi\nu}\left[-P^{\nu}_{\Lambda+\frac{1}{2}}(x)+
\cos\pi\nu\,\frac{\Gamma(\Lambda+\frac{3}{2}+\nu)}{\Gamma(\Lambda+\frac{3}{2}-\nu)}P^{-\nu}_{\Lambda+\frac{1}{2}}(x)\right]\,.
\end{equation}

\end{document}